\documentclass[aps,twocolumn,PRL,reprint,superscriptaddress]{revtex4}
\usepackage{amsmath}
\usepackage[percent]{overpic}
\usepackage{times}
\usepackage{subfigure}
\usepackage{latexsym}
\usepackage{graphicx}
\usepackage{bm}
\usepackage{amssymb}
\usepackage{anyfontsize}
\usepackage{hyperref}
\usepackage{multirow}
\usepackage{booktabs}
\usepackage{makecell}
\usepackage{natbib}
\usepackage{mathrsfs}
\usepackage{float}
\makeatletter

\newcommand{\Rmnum}[1]{\expandafter\@slowromancap\romannumeral #1@}
\makeatother

\begin{document}

\date{\today }

\title{Topological Invariant for Multi-Band Non-hermitian Systems with Chiral Symmetry}
\author{Chun-Chi Liu}
\affiliation{National Laboratory of Solid State Microstructures, Department of Physics, Nanjing University, Nanjing 210093, China}
\author{Liu-Hao Li}
\affiliation{National Laboratory of Solid State Microstructures, Department of Physics, Nanjing University, Nanjing 210093, China}
\author{Jin An}
\email{anjin@nju.edu.cn}
\affiliation{National Laboratory of Solid State Microstructures, Department of Physics, Nanjing University, Nanjing 210093, China}
\affiliation{Collaborative Innovation Center of Advanced Microstructures, Nanjing University, Nanjing 210093, China}

\begin{abstract}
  Topology plays an important role in non-hermitian systems. How to characterize a non-hermitian topological system under open-boundary conditions(OBCs) is a challenging problem. A one-dimensional(1D) topological invariant defined on a generalized Brillion zone(GBZ) was recently found to successfully describe the topological property of the two-band Su-Schrieffer-Heeger model. But for a 1D multi-band chiral symmetric system under OBCs, it is still controversial how to define the topological invariant. We show in this letter by exact proof and detailed demonstration that to acquire the topological invariant for multi-band non-hermitian models with chiral symmetry, the GBZ as the integral domain should be replaced by a more generalized closed loop. Our work thus establishes the non-Bloch bulk-boundary correspondence for 1D multi-band chiral symmetric non-hermitian systems.
\end{abstract}

\maketitle

\textbf{Introduction.}---
Recent studies on non-hermitian systems have revealed many new concepts and phenomena\cite{PhysRevB.56.8651,PhysRevB.58.8384,PhysRevLett.89.270401,PhysRevLett.101.150408,Rotter_2009,PhysRevLett.105.013903,PhysRevLett.113.250401,PhysRevX.4.041001,000423328000011,000455553900002,RevModPhys.93.015005,PhysRevB.106.115107,PhysRevLett.128.120401}, such as non-hermitian skin effect\cite{PhysRevLett.121.086803,PhysRevLett.123.170401,PhysRevLett.125.186802,PhysRevResearch.2.043167,PhysRevLett.124.086801,000588063600014,PhysRevLett.124.066602,PhysRevLett.127.116801,PhysRevB.103.045420,PhysRevB.103.L241408,PhysRevB.103.L140201,PhysRevB.106.235411,000791826000016,PhysRevLett.129.086601}, distinct differences between the open-boundary spectra(OBS) and the corresponding periodic-boundary spectra\cite{PhysRevLett.123.066404,PhysRevB.101.195147,PhysRevB.103.165123,PhysRevB.106.195425,PhysRevB.105.045422}, and coalesce of states, i.e., exception points\cite{PhysRevLett.86.787,PhysRevE.69.056216,000310465900017,zhen2015spawning,PhysRevLett.120.146402,000426366200036,PhysRevLett.123.066404,PhysRevLett.124.236403,PhysRevB.103.205205,PhysRevLett.126.086401,PhysRevLett.126.230402,PhysRevB.106.195425}. Topology is also found to play an increasingly important role in non-hermitian systems\cite{PhysRevB.84.153101,PhysRevB.84.205128,PhysRevA.87.012118,PhysRevLett.115.040402,PhysRevLett.116.133903,PhysRevLett.121.086803,PhysRevLett.121.136802,PhysRevLett.121.026808,PhysRevA.98.052116,PhysRevB.97.121401,PhysRevB.97.045106,PhysRevX.8.031079,PhysRevLett.120.146402,PhysRevA.97.052115,PhysRevX.9.041015,PhysRevLett.123.206404,PhysRevB.99.081103,PhysRevB.100.165430,PhysRevB.99.235112,PhysRevA.99.052118,PhysRevLett.123.246801,PhysRevB.99.245116,PhysRevB.100.035102,000519842400002,PhysRevLett.124.056802,PhysRevB.101.205417,PhysRevLett.126.216405,PhysRevLett.126.010401,PhysRevB.103.045420,RevModPhys.93.015005,PhysRevLett.126.216407,PhysRevB.103.075126,PhysRevB.105.075128,ding2022non,PhysRevB.105.094103,PhysRevB.106.195425,PhysRevLett.130.066601}. The central problem of non-hermitian topological systems is the non-Bloch bulk-boundary correspondence(BBC). Different from the hermitian counterparts\cite{classification1,RevModPhys.82.3045,RevModPhys.83.1057,RevModPhys.88.035005}, or non-hermitian systems with half-infinite boundary conditions, where the topological invariants can be defined on the Brillion Zone(BZ)\cite{PhysRevX.8.031079,PhysRevX.9.041015}, we have no fundamental principles to guide us to find a topological invariant to characterize a non-hermitian topological system under open-boundary conditions(OBCs). Even in one dimension, the non-Bloch BBC has not been well solved yet.

For 1D non-hermitian Su-Schrieffer-Heeger model, a pioneer work recently proposed that the topological invariant should be redefined on a generalized Brillion zone(GBZ) to capture the correct non-Bloch BBC\cite{PhysRevLett.121.086803}. The idea of replacement of BZ by GBZ has been further applied to two-dimensional systems and has been proven to be very successful\cite{PhysRevLett.121.136802,PhysRevB.105.075128,PhysRevB.107.035101}. This is feasible because in these models the GBZ is a unique closed loop or a unique closed surface. However, for 1D muti-band non-hermitian systems under OBCs, the situation changes completely. Generally, one has multiple subGBZs\cite{PhysRevLett.125.226402} and especially for systems with chiral symmetry the OBS consists of $N$ pairs of open arcs, and each pair is centrosymmetric about $E=0$ and corresponds to a closed loop(subGBZ). It is found that in some models, choosing one or some of the subGBZs to replace BZ in the definition of the topological invariant could not give the non-Bloch BBC. On the other hand, the number of topologically protected edge states(TPESs) changes even when the system is under a topologically trivial variation\cite{PhysRevLett.116.133903,PhysRevB.101.121116,PhysRevB.105.075420}. This means that the number of the TPESs cannot be the candidate of the invariant. Thus how to find out the topological invariant of the 1D multi-band non-hermitian systems with chiral symmetry becomes an open problem.

In this letter we solve this problem by redefining the topological invariant $\nu$ on a more generalized closed loop $\mathcal{L_{\beta}}$ in the complex $\beta$ plane and then by exactly proving the relevant statement. We show explicitly the number of the TPESs for systems characterized by $\nu$ can vary between $|\nu|$ and $2|\nu|$, and clarify the origin of the defectiveness of the TPESs. We discuss the implications of the conditions on the topological phase transitions imposed by our topological invariant.

\indent

\textbf{Topological invariant and the theorem.}---
In terms of $\beta=e^{ik}$, the model Hamiltonian of a 1D multi-band non-hermitian system can be expressed as
\begin{equation}
    H(\beta)=\sum_{m=-M_2}^{M_1}T_m \beta^{-m},
    \label{Hamiltonian}
\end{equation}
where $T_m$ is the $2N\times2N$ hoping matrix between the $m$th nearest-neighbor unit cells. The secular equation of $H(\beta)$ can be expressed as det$(H(\beta)-E)=\prod_{i=1}^{p+q}(\beta-\beta_{i}(E))/\beta^{p}=0$, where $p=2NM_{1}$, $q=2NM_{2}$. For a given energy $E$, it has $p+q$ $\beta$ roots, which are ordered in absolute value as $\vert\beta_{1}(E)\vert\leqslant ...\leqslant\vert\beta_{p+q}(E)\vert$. Each $\beta_{i}(E)$ corresponds to an eigenvector $\phi_{i}$, obeying the eigenequation,
\begin{equation}
    H(\beta_i(E))\phi_i=E\phi_i.
    \label{eigenhami}
\end{equation}
A state of energy $E$ belonging to the continuum OBS must obey $\vert\beta_p(E)\vert=\vert\beta_{p+1}(E)\vert$\cite{PhysRevLett.123.066404}. If the system holds chiral symmetry, then it can always take the following representation:
\begin{equation}
 H(\beta)=\left( \begin{array}{cc} 0 & R_{+}(\beta)\\R_{-}(\beta) & 0\\\end{array}\right),
 \label{formofhami}
\end{equation}
where $R_{\pm}$ is a $N\times N$ matrix. When the open system is gapful, the $E=0$ state would not belong to the continuum OBS, which means $\vert\beta_{p}(0)\vert\neq\vert\beta_{p+1}(0)\vert$. When the system is gapless, the $E=0$ state would be connected to the continuum OBS, with $\vert\beta_{p}(0)\vert=\vert\beta_{p+1}(0)\vert$. In order to study the topological property of this non-hermitian system, it's very meaningful to consider a class in which all the systems are topologically equivalent to it. Any two systems in the class under OBCs can be deformed continuously to each other without closing gap. We denote this topologically equivalent class as $\mathcal{C}_{\nu}$. Non-hermitian systems $H(\rho\beta)$ with $0<\rho<\infty$, which can be connected continuously by similarity transformations to $H(\beta)$, obviously belong to $\mathcal{C}_{\nu}$, since they share exactly the same OBS\cite{PhysRevB.105.045422}. However, different systems in the class may have different number of the TPESs. If $\mathcal{C}_{\nu}$ contains a hermitian system, then this system under OBCs is expected to have the maximum number of the TPESs in the class. For any closed loop $\mathcal{L_{\beta}}$ one can define the following winding number:
\begin{equation}
    \begin{split}
        &\nu=\frac{1}{2\pi i}\oint_{\mathcal{L_{\beta}}}{\rm Tr}(q^{-1}dq)\\
        &=\frac{1}{4\pi i}\oint_{\mathcal{L_{\beta}}}d\beta \frac{d}{d\beta}({\rm ln}\;{\rm det}(R_{+}(\beta))-{\rm ln}\;{\rm det}(R_{-}(\beta))),\\
    \end{split}
    \label{winding}
\end{equation}
where $q$ comes from the standard occupation projection $Q$ operator\cite{classification1}. Our theorem is: \textit{if the closed loop $\mathcal{L_{\beta}}$ encircles the first $p$ $\beta$ roots of $E=0$ with the other $q$ $\beta$ roots being kept outside, the winding number $\nu$ gives the topological invariant that characterizes the topological feature of the 1D chiral-symmetric non-hermitian systems under open boundary conditions, i.e., $|\nu|$ is equal to the minimum number of the TPESs of the systems in the topologically equivalent class $\mathcal{C}_{\nu}$ and the number of the TPESs of any systems in $\mathcal{C}_{\nu}$ is no more than $2|\nu|$.} Obviously, $\mathcal{L_{\beta}}$ is not unique. A useful and convenient selection of $\mathcal{L_{\beta}}$, particular in numerical computation of the invariant, is a perfect circle centered at the origin $\beta=0$ with radius $\rho$ obeying $\vert\beta_{p}(0)\vert<\rho<\vert\beta_{p+1}(0)\vert$. On the other hand, if one of the subGBZs just contains the first $p$ $\beta$ roots, it can also serve as a possible $\mathcal{L_{\beta}}$.

\indent
$N=1$ is a special case which corresponds to the two-band models and has been discussed previously in Refs\cite{PhysRevLett.121.086803}, where the GBZ $C_{\beta}$ is proposed as the closed loop in the integral of the winding number. This is consistent with our theorem, since the GBZ has been proven to be the boundary between $\beta_{p}(E)$ and $\beta_{p+1}(E)$ regions for all $E$\cite{PhysRevLett.125.126402}. So for any state of $E$ not on the continuum OBS, the GBZ exactly encircles only the first $p$ $\beta$ roots of $E$, so the GBZ is actually one of the possible $\mathcal{L_{\beta}}$ defined here. However, when $N\geq 2$, the situation becomes much more complicated since the system generally has $2N$ bands and the GBZ is composed of $N$ closed loops(subGBZs). One or several of the subGBZs, or even all the subGBZs may be chosen as the integral loop(s), but they are all found to be questionable. Therefore, our theorem can be viewed as solving this problem in the case of multi-band chiral symmetric non-hermitian systems by choosing $\mathcal{L_{\beta}}$ instead of the GBZ as the integral loop in the definition of the topological invariant.

\begin{figure}
    \centering
    \setlength{\belowcaptionskip}{-0.1cm}
    \includegraphics[width=1\columnwidth]{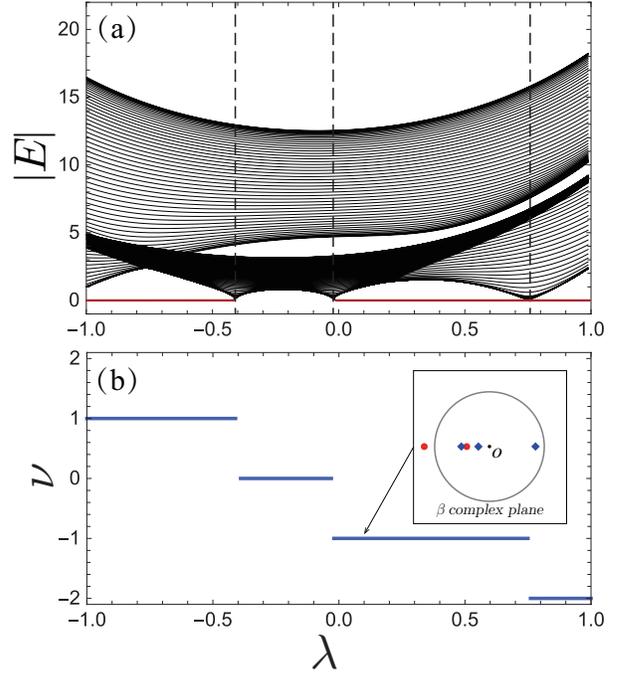}
    \caption{An example exhibiting the consistence between the occurrence of zero modes and our topological invariant. (a): Absolute value of energy of the states under OBCs as a function of parameter $\lambda$, where the thick lines denotes the topological zero modes. (b): Topological invariant $\nu$ versus $\lambda$. Here $R_{\pm}=t_0^{\pm}+t_x^{\pm}\sigma_x+t_y^{\pm}\sigma_y+t_z^{\pm}\sigma_z$, where $t_0^+=4+1.5\beta^{+}$, $t_x^+=4+2.5/\beta^{+}$, $t_y^+=-i+0.5i/\beta^{+}$, $t_z^+=3+0.5\beta^{+}$ and $t_0^-=4.5+1.5/\beta^{-}$, $t_x^-=3+2.5\beta^{-}$, $t_y^-=-i+0.5i\beta^{-}$, $t_z^-=2.5+0.5/\beta^{-}$ with $\beta^{\pm}=4^{\mp\lambda}\beta$. The inset gives the perfect circle in $\beta$ plane used to calculate the invariant at a certain $\lambda$, with the solid circles(diamonds) being $\beta$ roots of $E=0$ from det$R_{+}=0$(det$R_{-}=0$).
        }
    \label{fig1}
\end{figure}

\indent
We take a four-band($N=2$) model as an example to illustrate the theorem. The variations of the OBS with parameter $\lambda$ are shown in Fig.\ref{fig1}(a), while the topological invariant $\nu$ calculated by Eq.(\ref{winding}) is shown in Fig.\ref{fig1}(b). The nontrivial regions given by $\nu$ agree very well with that predicted by the direct numerical diagonalization of the open chain of the system.\\
\indent
\textbf{Proof of the theorem.}---
Here we give the proof of our theorem. We will derive the solutions of the isolated $E=0$ state under OBCs and then connect its degeneracy to the topological invariant $\nu$ we defined. For a gapful system, $E=0$ state does not belong to the continuum OBS, which means $\vert \beta_p(0) \vert\neq\vert \beta_{p+1}(0) \vert$. We start from a system with property $\vert \beta_p(0) \vert<1<\vert \beta_{p+1}(0) \vert$. If the non-hermitian system $H(\beta)$ we study does not have this property, one can always choose a $H(\rho\beta)$ system as the starting one, since its $\beta$ roots of $E=0$ are just those of $H(\beta)$, but scaled by a factor $\rho^{-1}$. Thus for a 1D open chain with $L$ unit cells, the wave function of a state with energy $E=0$ can be expanded as the superposition of its $p+q$ eigenmodes $\phi_{i}$:
\begin{equation}
        \psi(j)=\sum_{i=1}^{p}c_i(\beta_i(0))^j \phi_i +\sum_{i=1}^{q}c_{i+p}(\beta_{p+i}(0))^{j-L}\phi_{p+i},
        \label{wavefun}
\end{equation}
where $j$ represents the $j$th unit cell, and $c_i$ is the superposition coefficients of the eigenmodes $\phi_i$.

Now we turn to the boundary conditions. We consider a simplified case where both $T_{M_{1}}$ and $T_{-M_{2}}$ are invertible. If this is not the case in practice, for a gapful system, one can slightly change them to be invertible without closing the gap and so without changing the topological property of the system. By introducing some extra unit cells near the left and right boundaries, the OBCs become\cite{sup}:
\begin{equation}
   \begin{split}
     &\psi(0)=\psi(-1)=...=\psi(-M_1 +1)=0,\\
     &\psi(L+1)=\psi(L+2)=...=\psi(L+M_2)=0.
     \label{obc}
   \end{split}
\end{equation}
Below we first study the case in the thermodynamic limit and then we take the effect of a finite and sufficiently large $L$ into account. As $L\rightarrow\infty$, the OBCs can be expressed as:
\begin{equation}
    \begin{split}
        &\sum_{i=1}^{p}c_i(\beta_i(0))^j\phi_i=0, \quad\quad j=0,-1,...,-M_1+1.\\
        &\sum_{i=1}^{q}c_{i+p}(\beta_{i+p}(0))^{j}\phi_{i+p}=0, \quad\quad j=1,2,...,M_2.
    \end{split}
    \label{plpr}
\end{equation}
These two sets of equations can be reexpressed as $\mathcal{M_L}\vert \mathcal{L}\rangle=0$ and $\mathcal{M_R}\vert \mathcal{R}\rangle=0$, where $\vert \mathcal{L}\rangle=(c_1,c_2,...,c_p)^T$ and $\vert \mathcal{R}\rangle=(c_{p+1},c_{p+2},...,c_{p+q})^T$ respectively, and the $p\times p$ matrix $\mathcal{M_L}$ and $q\times q$ matrix $\mathcal{M_R}$ are given below,
\begin{equation}
    \begin{split}
        &\mathcal{M_L}=(\Phi_{1,\mathcal{L}}, \Phi_{2,\mathcal{L}},..., \Phi_{p,\mathcal{L}}),\\
        &\mathcal{M_R}=(\Phi_{p+1,\mathcal{R}}, \Phi_{p+2,\mathcal{R}},..., \Phi_{p+q,\mathcal{R}}),
    \end{split}
    \label{mlmr}
\end{equation}
with $\Phi_{i,\mathcal{L}}\equiv(\phi_i,\beta_i^{-1}\phi_i,...,\beta_i^{-M_1+1}\phi_i)^T$, $\Phi_{i,\mathcal{R}}\equiv(\beta_i\phi_i,\beta_i^{2}\phi_i,...,\beta_i^{M_2}\phi_i)^T$. Whether $\mathcal{M_L}$ or $\mathcal{M_R}$ is singular determines directly whether there exists $E=0$ solution.

The detailed form of $\mathcal{M_L}$ or $\mathcal{M_R}$ relies on that of $H(\beta)$ in the presence of chiral symmetry, Eq.(\ref{formofhami}). Let $\phi_i=(u_i,v_i)^T$, the eigenequation for $E=0$ becomes:
\begin{equation}
    \begin{split}
        &R_+(\beta_i(0))v_i=0,\\
        &R_-(\beta_i(0))u_i=0.\\
    \end{split}
    \label{viui}
\end{equation}
If $\beta_i(0)$ comes from the the $k$th root of $\rm{det}R_+=0$, then $(\beta_i(0),\phi_i)$ is also denoted as $(\beta^{+}_k(0),\phi^{+}_k)$ with $\phi_i=(0,v_i)^T$. Otherwise $\beta_i(0)$ may come from the $k$th root of $\rm{det}R_-=0$, so $(\beta_i(0),\phi_i)$ is also denoted as $(\beta^{-}_k(0),\phi^{-}_k)$ with $\phi_i=(u_i,0)^T$. Therefore, among the $p+q$ $\beta$ roots, half is from $\rm{det}R_+=0$ and half from $\rm{det}R_{-}=0$. In the first(last) $p(q)$ roots we assume there are $p_1(q_{1})$ ones from $\rm{det}R_+=0$ and $p_2(q_2)$ ones from $\rm{det}R_-=0$. Naturally we have $p_1+p_2=p$, $q_1+q_2=q$ and $p_1+q_1=p_2+q_2=(p+q)/2$. Among $p_{1}$, $p_{2}$, $q_{1}$, $q_{2}$, only one is independent and so by introducing an integer $w$ they can be parameterized as follows:
\begin{equation}
  \begin{split}
  & p_{1}=p/2+w, \quad p_{2}=p/2-w,\\
  & q_{1}=q/2-w, \quad q_{2}=q/2+w.
  \end{split}
  \label{pq}
\end{equation}
Our topological invariant $\nu$ defined in Eq.(\ref{winding}) can then be given by:
\begin{equation}
\nu=(p_{1}-p_{2})/2=w.
\label{w}
\end{equation}
So we have to prove a state of system denoted by $w$ must possess at least $\vert w\vert$ TPESs. Let $\Phi_{i,\mathcal{L}}^{\pm}\equiv(\phi_i^{\pm},(\beta_i^{\pm})^{-1}\phi_i^{\pm},...,(\beta_i^{\pm})^{-M_1+1}\phi_i^{\pm})^T$ and $\Phi_{i,\mathcal{R}}^{\pm}\equiv(\beta_i^{\pm}\phi_i^{\pm},(\beta_i^{\pm})^{2}\phi_i^{\pm},...,(\beta_i^{\pm})^{M_2}\phi_i^{\pm})^T$ and by performing some matrix column exchanges on $\mathcal{M_L}$ or $\mathcal{M_R}$, the two matrices would be equivalently changed to be:
\begin{equation}
    \begin{split}
        &\mathcal{M_L}\rightarrow(\Phi_{1,\mathcal{L}}^+,...,\Phi_{p_1,\mathcal{L}}^+,\Phi_{1,\mathcal{L}}^-,...,\Phi_{p_2,\mathcal{L}}^-)
        \equiv(\Phi_{\mathcal{L}}^+,\Phi_{\mathcal{L}}^-),\\
        &\mathcal{M_R}\rightarrow(\Phi_{p_1+1,\mathcal{R}}^+,...,\Phi_{p_1+q_1,\mathcal{R}}^+,\Phi_{p_2+1,\mathcal{R}}^-,...,\Phi_{p_2+q_2,\mathcal{R}}^-)\\
        &\equiv(\Phi_{\mathcal{R}}^+,\Phi_{\mathcal{R}}^-).
    \end{split}
    \label{ro-mlmr}
\end{equation}
with $\vert\mathcal{L}\rangle$ and $\vert\mathcal{R}\rangle$ also being changed to be $\vert\mathcal{L}\rangle \rightarrow(c^+_1,...,c^+_{p_1},c^-_1,...,c^-_{p_2})^T\equiv(C^+,C^-)^T$, and $\vert\mathcal{R}\rangle \rightarrow(c^+_{p_1+1},...,c^+_{p_1+q_1},c^-_{p_2+1},...,c^-_{p_2+q_2})^T\equiv(D^+,D^-)^T$.

\indent
One particular feature to be noted is that if det$\mathcal{M_L}\neq0$, we must have $p_{1}=p_{2}=p/2$, namely, $w=0$. To prove this statement, we assume that for example, $p_{1}>p_{2}$. Since for any $\Phi_{i,L}^{+}$-like vector with $p$ components, one half of its components is zero due to $\phi_{i}=(0,v_{i})^T$, the dimension of the space spanned by this kind of vectors is at most $p/2$. So the rank of $\mathcal{M_L}$ is at most $p/2+p_{2}<p$, leading to the contradiction that det$\mathcal{M_L}=0$. Similarly, one can prove that if det$\mathcal{M_R}\neq0$, one must have $w=0$.

\indent
We note that the solutions of $\mathcal{M_L}\vert \mathcal{L}\rangle=0$ or $\mathcal{M_R}\vert \mathcal{R}\rangle=0$ are actually solutions of $E=0$ state for the half-infinite system with left-open boundary or right-open boundary. But different systems in the same topologically equivalent class may have different number of solutions, due to the existence of non-topological edge states. To exclude these edge states not protected by chiral symmetry, the number of solutions must be minimized by continuously deforming the system without closing the gap. In supplementary material\cite{sup} we demonstrate another generic important feature that by keeping the open system gapful, one can continuously vary $R_{\pm}(\beta)$ by continuously moving the $\beta$ roots of $E=0$, to make Rank$\mathcal{M_L}$ and Rank$\mathcal{M_R}$ maximized. Therefore, for any state of system under OBCs, it can always be made to be a topologically equivalent state with Rank$\mathcal{M_L}$ and Rank$\mathcal{M_R}$ being maximized. After this rank maximization process, for a state denoted by $w$, we have Rank$\mathcal{M_L}=p/2+(p/2-\vert w\vert)=p-\vert w\vert$, and Rank$\mathcal{M_R}=q/2+(q/2-\vert w\vert)=q-\vert w\vert$. According to linear algebra theory, both $\mathcal{M_L}\vert \mathcal{L}\rangle=0$ and $\mathcal{M_R}\vert \mathcal{R}\rangle=0$ have $\vert w\vert$ independent solutions. These solutions corresponds to the $|w|$ left-localized TPESs or $\vert w\vert$ right-localized TPESs when the system is under the left-open or right-open half-infinite boundary conditions. Thus the rank maximization process can be understood as a modification of the system which eliminates the solutions which are not protected by chiral symmetry, since the original state of system denoted by $w$ may have more solutions if the ranks of $\mathcal{M_L}$ and $\mathcal{M_R}$ are not being maximized.

We further show that when the finiteness of $L$ is taken into account, $\vert w\vert$ is the minimal number of the TPESs for systems in the topological equivalent class. A large but finite $L$ indicates the boundary equations in Eq.(\ref{mlmr}) should be replaced by
\begin{equation}
    \begin{split}
        &\left(
            \begin{array}{cc}
                \mathcal{M_L} &\mathcal{M_{LR}}\\
                \mathcal{M_{RL}} &\mathcal{M_{R}}
            \end{array}
        \right)\vert \psi  \rangle=0,\\
        &\vert \psi  \rangle=(c_1,...,c_p,c_{p+1},...,c_{p+q})^T,
    \end{split}
    \label{fullm}
\end{equation}
with $\mathcal{M_{LR}}=(\beta_{p+1}^{-L}\Phi_{p+1,\mathcal{L}},..., \beta_{p+q}^{-L}\Phi_{p+q,\mathcal{L}})$, and $\mathcal{M_{RL}}=(\beta_{1}^{L+1}\Phi_{1,\mathcal{R}},..., \beta_{p}^{L+1}\Phi_{p,\mathcal{R}})$, followed by the same column exchanges as $\mathcal{M_{L}}$ and $\mathcal{M_{R}}$ in Eq.(\ref{ro-mlmr}). For concreteness, we assume $w\geq0$. Because Rank$\Phi_{\mathcal{L}}^-=p/2-w$, Rank$\Phi_{\mathcal{R}}^+=q/2-w$, one can easily show that $\mathcal{M_L}\vert \mathcal{L}\rangle=0$($\mathcal{M_R}\vert \mathcal{R}\rangle=0$) would indicate $C^{-}=0(D^{+}=0)$. This implies that the left-localized states reside on sublattice $B$, while the right-localized states reside on sublattice $A$\cite{sup1}. This is actually required by chiral symmetry. We denote the solutions as $\vert \mathcal{L}_{m}\rangle=(C^{+}_{m},0)^{T}$($\vert \mathcal{R}_n\rangle=(0,D^{-}_{n})^{T}$), $m(n)=$1, 2,...,$w$. One can construct $w\times w$ matrix $M_{LR/RL}$, whose $mn$ matrix entry is $\langle\mathcal{(L/R)}_{m}\vert\mathcal{M_{LR/RL}}\vert\mathcal{(R/L)}_{n}\rangle$ and is generally proportional to $(\beta^{-}_k(0))^{-L}/$or $(\beta^{+}_l(0))^{L}$, with $k\geq p_{2}+1=p/2-w+1$ and $l\leq p_{1}=p/2+w$. So according to the degenerate perturbation theory, the actual solutions for the TPESs for a finite $L$ are $\vert\psi^{\pm}_{m}\rangle=(\mathcal{A}_{m}\vert \mathcal{L^{\prime}}_{m}\rangle,\pm \mathcal{B}_{m}\vert \mathcal{R^{\prime}}_{m}\rangle)^T$, where $m=$1, 2,...,$w$. Here $\vert \mathcal{L^{\prime}}_{m}\rangle$ and $\vert \mathcal{R^{\prime}}_{m}\rangle$ are the linear superpositions of $\vert \mathcal{L}_{m}\rangle$ and $\vert \mathcal{R}_{m}\rangle$ respectively. $\vert\psi^{+}_{m}\rangle$ is related to $\vert\psi^{-}_{m}\rangle$ by chiral symmetry: $\mathcal{S}\vert\psi^{\pm}_{m}\rangle=-\vert\psi^{\mp}_{m}\rangle$, with $\mathcal{S}$ the chiral symmetry operator. $\mathcal{A}_{m}$ and $\mathcal{B}_{m}$ are the coefficients, and their ratios are found to be:
\begin{equation}
    \frac{\mathcal{B}_{m}}{\mathcal{A}_{m}}=\frac{\sqrt{\langle \mathcal{R^{\prime}}_{m}\vert\mathcal{M_{RL}}\vert \mathcal{L^{\prime}}_{m} \rangle}}{\sqrt{\langle \mathcal{L^{\prime}}_{m}\vert\mathcal{M_{LR}}\vert \mathcal{R^{\prime}}_{m} \rangle}},
    \label{a/b}
\end{equation}
which is proportional to $(\beta^{+}_l(0)\beta^{-}_k(0))^{L/2}$. So for a relatively small $L$, among the $2NL$ states of the open chain, $2w$ ones belong to the isolated $E=0$, while the other ones belong to the continuum OBS. However, the ratios are very sensitive to the details of the system, and as $L$ is sufficiently large, they can be finite, but can also be vanishingly small or be approaching infinity. If a ratio is not finite, the corresponding pair of the TPESs can be viewed as being coalescing into one single state so the final number of the TPESs is actually varying from $w$ to $2w$ for systems in the same topologically equivalent class. This is the origin of the defectiveness of the TPESs. For hermitian systems, all the ratios are expected to be finite, and the hermitian open chains always have 2$w$ TPESs localized at both ends. In the same topologically equivalent class, there always exist some systems which have the minimum number of the TPESs: $w$. Actually, the systems described by $H(\rho\beta)$ possess $w$ TPESs, as long as $\rho$ is sufficiently small or large. This is because due to the similarity transformation, all superposition coefficients for systems of $H(\rho\beta)$ would be scaled by a factor $\rho^{-L}$, compared to $H(\beta)$. Therefore, for a sufficiently large(small) $\rho$, the open systems described by $H(\rho\beta)$ have exactly $\vert w\vert$ left-(right-)localized TPESs which all reside on sublattice B(A) if $w>0$, or on sublattice A(B) if $w<0$. This means the number of the TPESs is not a topological invariant, and $\vert w\vert$ or $\vert\nu\vert$ is proved to be the minimum number of the TPESs and we then complete the proof.

\begin{figure}
    \setlength{\belowcaptionskip}{-0.1cm}
    \centering
    \includegraphics[width=1.0\columnwidth]{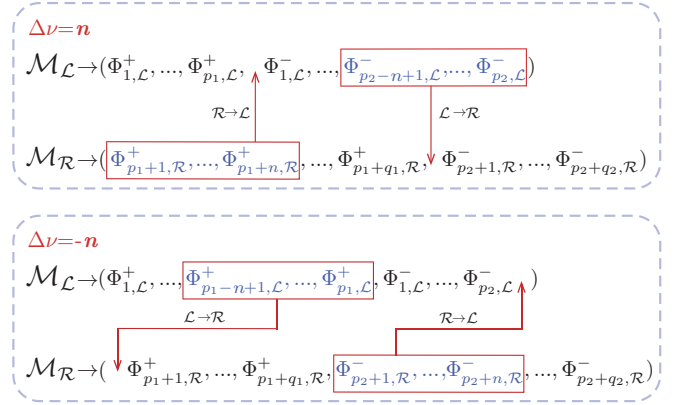}
    \caption{Schematic demonstration of topological transitions with $\Delta\nu=\pm n$, where $n$ $\beta$ roots of det$R_{+}=0$ and $n$ roots of det$R_{-}=0$ are exchanged, obeying $\vert\beta_{p-n+1}(0)\vert=...=\vert\beta_{p}(0)\vert=\vert\beta_{p+1}(0)\vert=...=\vert\beta_{p+n}(0)\vert$.}
    \label{fig2}
\end{figure}
\indent
\textbf{Topological transitions.}---
By varying parameters, the system would undergo topological transitions when the invariant integer $\nu$ changes, at which the system becomes gapless, indicating $\vert\beta_{p}(0)\vert=\vert\beta_{p+1}(0)\vert$. At this critical point, if $\Delta\nu=\pm1$, the state of system denoted by $w$ would be changed to be one denoted by $w\pm1$, implying $\vert\beta^{+}_{p/2+\vert w\vert}(0)\vert=\vert\beta^{-}_{p/2-\vert w\vert+1}(0)\vert$ or $\vert\beta^{-}_{p/2+\vert w\vert}(0)\vert=\vert\beta^{+}_{p/2-\vert w\vert+1}(0)\vert$. So among $\beta_{p}(0)$ and $ \beta_{p+1}(0)$, one is from ${\rm det}R_{+}=0$, the other is from det$R_{-}=0$  and their exchange means topological transition. We remark that when both $\beta_{p}(0)$ and $ \beta_{p+1}(0)$ come from ${\rm det}R_{+}=0$ or ${\rm det}R_{-}=0$, their exchanges would not induce topological transitions, but may drive the system to a topologically equivalent state or a gapless state with exceptional points\cite{PhysRevLett.123.066404}. If $\Delta\nu=\pm2$, similar discussion would lead to the conclusion that at the transition point, we must have $\vert\beta_{p-1}(0)\vert=\vert\beta_{p}(0)\vert=\vert\beta_{p+1}(0)\vert=\vert\beta_{p+2}(0)\vert$, and among the four $\beta$ roots, two must come from det$R_{+}=0$, with the other two from det$R_{-}=0$. We show schematically the general case of $\Delta\nu=\pm n$ in Fig.(\ref{fig2}). Now we make a few remarks on the relationship between subGBZs and topological transition. At the transition point, the $E=0$ state is connected to one energy branch of the OBS, which corresponds to a definite subGBZ. In supplementary material\cite{sup} we prove that near the phase transition, if this subGBZ is chosen as the closed loop $\mathcal{L_{\beta}}$ in the integral in Eq.(\ref{winding}), the change of this newly defined winding number $\nu^{'}$ at the transition point with $\vert\beta_{p}(0)\vert=\vert\beta_{p+1}(0)\vert$ would be $\Delta\nu^{'}=\pm1$.

\indent
\textbf{Further discussion and conclusion.}---
The phase winding of ${\rm det}H$ along the closed loop $\mathcal{L_{\beta}}$ is the sum of those of ${\rm det}R_{+}$ and ${\rm det}R_{-}$, and can be easily checked to be zero. So the topological invariant $\nu$ can either be expressed as the phase winding of ${\rm det}R_{+}$ or minus that of ${\rm det}R_{-}$ along $\mathcal{L_{\beta}}$. This does not mean that either $R_{+}$ or $R_{-}$ can independently determine the invariant because $\mathcal{L_{\beta}}$ has already contained the information about both of them. For any non-hermitian chiral symmetric open system described by $H(\beta)$ with $\vert\beta_{p}(0)\vert\neq\vert\beta_{p+1}(0)\vert$, the invariant $\nu$ can be computed directly from Eq.(4), but can also be understood physically as follows. Choose its topologically equivalent system $H(\rho\beta)$ obeying $\vert\beta_{p}(0)\vert/\rho<1<\vert\beta_{p+1}(0)\vert/\rho$. This new system takes the same invariant $\nu$ but its $\mathcal{L_{\beta}}$ can be chosen to the unit circle, implying $\nu$ can now be defined on the BZ for this system. Bulk invariants defined on the BZ in non-hermitian systems means that the invariants are coming from the systems under periodic boundary conditions, and the corresponding BBC connects them to the TPESs of the corresponding half-infinite systems. Our topological invariant characterizes the non-Bloch BBC which connects the invariant to the TPESs under OBCs. So for this system, the topological invariants for the conventional non-hermitian BBC and non-Bloch BBC become identical. This may shed light on the relationship between BBC and non-Bloch BBC in high-dimensional non-hermitian topological systems.\\

\indent
This work is supported by NSFC under Grants No.11874202.
\bibliographystyle{unsrt}
\bibliography{ref}

\begin{thebibliography}{10}
\bibitem{sup1}
Divide the open chain into two parts: sublattice a and sublattice b, which
  correspond to $u$ and $v$ components in $\phi$, respectively.
\end{thebibliography}
\end{document}


\date{\today }

\title{Supplemental materials for ``Topological Invariant for Multi-Band Non-hermitian Systems with Chiral Symmetry''}
\author{Chun-Chi Liu}
\affiliation{National Laboratory of Solid State Microstructures, Department of Physics, Nanjing University, Nanjing 210093, China}
\author{Liu-Hao Li}
\affiliation{National Laboratory of Solid State Microstructures, Department of Physics, Nanjing University, Nanjing 210093, China}
\author{Jin An}
\email{anjin@nju.edu.cn}
\affiliation{National Laboratory of Solid State Microstructures, Department of Physics, Nanjing University, Nanjing 210093, China}
\affiliation{Collaborative Innovation Center of Advanced Microstructures, Nanjing University, Nanjing 210093, China}

\maketitle

\centerline{\textbf{\Rmnum{1}.Open Boundary Conditions(OBCs) and Their Simplification}}

\quad

The eigenequation of $H(\beta)$ can be written as:
\begin{equation}
    \sum_{m=-M_2}^{M_1}T_{m}\psi(j-m)=E\psi(j),
    \label{eq11}
\end{equation}
where $M_{1}<j<L-M_{2}$. When $1\leq j\leq M_{1}$, the eigenequation becomes the left-boundary conditions, which are given by:

\begin{equation}
    \begin{split}
        &\sum_{m=-M_2}^{M_1-1}T_{m}\psi(M_1-m)=E\psi(M_1),\\
        &\sum_{m=-M_2}^{M_1-2}T_{m}\psi(M_1-1-m)=E\psi(M_1-1),\\
        &\quad\quad\vdots\\
        &\sum_{m=-M_2}^{0}T_{m}\psi(1-m)=E\psi(1).\\
    \end{split}
    \label{eq12}
\end{equation}
When $L-M_{2}+1\leq j\leq L$, the eigenequation becomes the right-boundary conditions, which are:
\begin{equation}
    \begin{split}
        &\sum_{m=-M_2+1}^{M_1}T_{m}\psi(L-M_2+1-m)=E\psi(L-M_2+1),\\
        &\sum_{m=-M_2+2}^{M_1}T_{m}\psi(L-M_2+2-m)=E\psi(L-M_2+2),\\
        &\quad\quad\vdots\\
        &\quad\sum_{m=0}^{M_1}\quad T_{m}\psi(L-m)=E\psi(L).\\
    \end{split}
    \label{eq13}
\end{equation}
\begin{figure}
    \centering
    \includegraphics[width=0.8\textwidth]{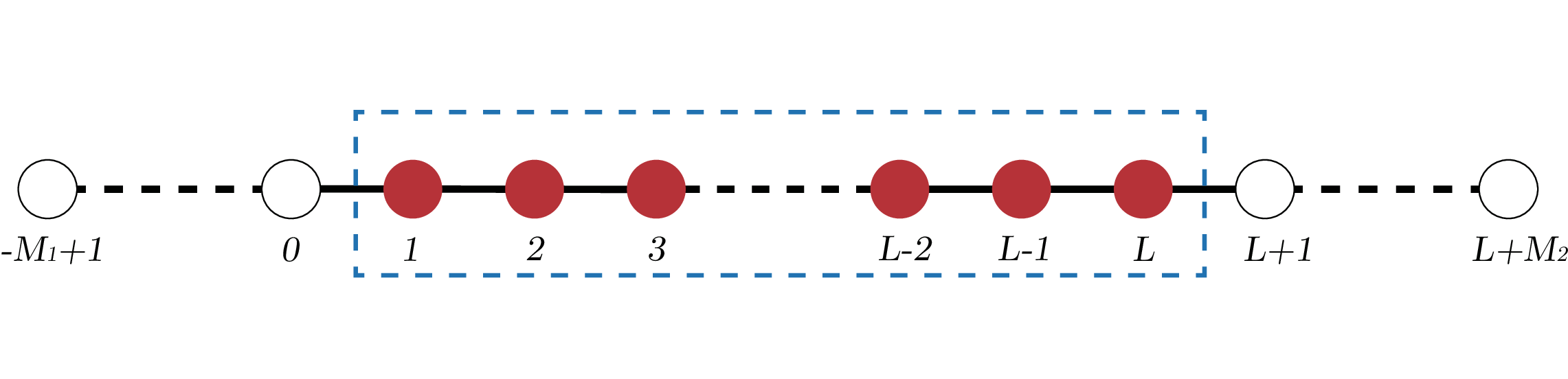}
    \caption{A one-dimensional open chain with $L$ unit cells represented by solid circles, where each unit cell contains $2N$ internal degrees of freedom. Here the hollow circles at both ends of the chain are the extra unit cells added for the discussion of open-boundary conditions.}
    \label{sfig1}
\end{figure}
The two sets of equations contain $M_1$ and $M_2$ equations respectively. In order to make the eigenequation Eq.(\ref{eq11}) be obeyed for all unit cells, including those near the left and right boundaries, one can introduce extra unit cells as shown in Fig.1. Combination of these new equations and the above two sets of equations lead to the following simplified left-boundary conditions:
\begin{equation}
    \begin{split}
        &T_{M_1}\psi(0)=0,\\
        &T_{M_1}\psi(-1)+T_{M_1-1}\psi(0)=0,\\
        &\quad \vdots\\
        &T_{M_1}\psi(-M_1+1)+T_{M_1-1}\psi(-M_1+2)+...+T_{1}\psi(0)=0,
    \end{split}
    \label{eq14}
\end{equation}
and right-boundary conditions:
\begin{equation}
    \begin{split}
        &T_{-M_2}\psi(L+1)=0,\\
        &T_{-M_2}\psi(L+2)+T_{-M_2+1}\psi(L+1)=0,\\
        &\quad \vdots\\
        &T_{-M_2}\psi(L+M_2)+T_{-M_2+1}\psi(L+M_2-1)+...+T_{-1}\psi(L+1)=0.
    \end{split}
    \label{eq15}
\end{equation}
If both $T_{M_{1}}$ and $T_{-M_{2}}$ are invertible, then according to the first equations of the two sets of equations, we have: $\psi(0)=\psi(L+1)=0$. From the second equations, we further have $\psi(-1)=\psi(L+2)=0$. Along the similar argument, we would find the following equivalent boundary equations:
\begin{equation}
    \begin{split}
        &\psi(0)=\psi(-1)=...=\psi(-M_1+1)=0,\\
        &\psi(L+1)=\psi(L+2)=...=\psi(L+M_2)=0.\\
    \end{split}
    \label{eq16}
\end{equation}
These boundary conditions can be understood as that the wave function $\psi(j)$ at the extra unit cells denoted by the hollow circles in Fig.\ref{sfig1} should all be zero. Let's consider the system described by $H(\rho\beta)$ whose wave function is denoted by $\widetilde{\psi}(j)$. The real-space Hamiltonian $\mathcal{H}\rho$ of $H(\rho\beta)$ and the real-space Hamiltonian $\mathcal{H}$ of $H(\beta)$ are related by a similarity transformation $\mathcal{H}\rho=S^{-1}\mathcal{H}S$, with $S={\rm diag}(\rho^{1},\rho^{2},...,\rho^{L})\otimes I_{2N}$. Accordingly, we have $\widetilde{\psi}(j)=\rho^{-j}\psi(j)$ and the hopping matrices of $H(\rho\beta)$ become $T_{m}\rho^{-m}$, so for $\widetilde{\psi}(j)$ we can obtain the similar OBCs to Eq.(\ref{eq16}).

\quad

\quad

\centerline{\textbf{\Rmnum{2}.Rank Maximization of Matrices $\mathcal{M_L}$ and $\mathcal{M_R}$}}

\quad

Introduce a continuous parameter $\Lambda$, and $R_{\pm}$ can be viewed as functions of $\Lambda$: $R_{\pm}(\beta,\Lambda)$ obeying $R_{\pm}(\beta,0)=R_{\pm}(\beta)$. Here we shall prove in this section that for a gapful open system with $\vert\beta_{p}(0)\vert\neq\vert\beta_{p+1}(0)\vert$, while keeping the system gapful, one can always vary the parameter $\Lambda$ to continuously modify the positions of the $\beta$ roots of det$R_{\pm}(\beta,\Lambda)=0$ to make the rank of matrices of $\mathcal{M_L}$ and $\mathcal{M_R}$ in the main text maximized, without breaking the chiral symmetry. This kind of variation of the system is obviously a topologically equivalent one, so the final state should share the same topological invariant $\nu$ with the initial one.

By solving the eigenequation of $E=0$:
\begin{equation}
    H(\beta)\phi_i=0\rightarrow
 \begin{cases}
    R_+(\beta)v_i=0\\
    R_-(\beta)u_i=0,
\end{cases}
\label{eq21}
\end{equation}
we get the $\beta$ roots and their corresponding eigenvectors: $\beta^{+}_{i}$, $\phi_i^+=(0,v_i)^T$($\beta^{-}_{i}$, $\phi_i^-=(u_i,0)^T$), $i=1$, $2$, ..., $N(M_1+M_2)$. From all the eigenvectors $v_i$ of $R_{+}$, we construct a $N(M_1+M_2)\times N(M_1+M_2)$ matrix $\mathcal{M}_{N}^+$:
\begin{equation}
    \mathcal{M}_{N}^{+}=(\Phi^+_1,\Phi^+_2,...,\Phi^+_{N(M_1+M_2)}),
    \label{eq22}
\end{equation}
with the column vector $\Phi^+_i=(v_i,(\beta_i^+)^{-1}v_i,...,(\beta_i^+)^{-M_1-M_2+1}v_i)^T=(1,(\beta_i^+)^{-1},...,(\beta_i^+)^{-M_1-M_2+1})^T\otimes v_i$.

When $N=1$ or in the two-band case, the matrix $\mathcal{M}_{1}^{+}$ becomes:
\begin{equation}
    \begin{split}
        \mathcal{M}_{1}^{+}=\left(
            \begin{array}{cccc}
             1&1&...&1
               \\
             (\beta_1^+)^{-1} & (\beta_2^+)^{-1} & ... & (\beta_{M_1+M_2}^+)^{-1} \\
             \vdots & \vdots & \vdots & \vdots \\
             (\beta_1^+)^{-M_1-M_2+1} & (\beta_2^+)^{-M_1-M_2+1} & ... & (\beta_{M_1+M_2}^+)^{-M_1-M_2+1} \\
            \end{array}
            \right),
    \end{split}
    \label{eq23}
\end{equation}
which is a Vandermonde matrix, whose determinant is a well-known polynomial $\mathcal{P}_{V}$ of $(\beta_i^+)^{-1}$:
\begin{equation}
\text{det}\mathcal{M}_{1}^{+}=\prod_{1\leq i<j\leq M_1+M_2}((\beta_i^+)^{-1}-(\beta_j^+)^{-1})\equiv \mathcal{P}_{V}((\beta_1^+)^{-1},(\beta_2^+)^{-1},...,(\beta^{+}_{M_{1}+M_{2}})^{-1}),
\label{eq24}
\end{equation}
If all $\beta_i^+$ are different from each other, then obviously det$\mathcal{M}_{1}^{+}\neq0$. Otherwise, there exist repeated $\beta$ roots, then by fixing the other roots we can easily do a parameter change to make the repeated roots split without closing the gap, since the two roots $\beta_{p}(0)$ and $\beta_{p+1}(0)$ with unequal magnitude are still left unchanged. So for any state of the system, either this state or its topologically equivalent one would always obey: det$\mathcal{M}_{1}^{+}\neq0$. For later convenience, we call vectors like $(1,(\beta_i^+)^{-1},...,(\beta_i^+)^{-M_1-M_2+1})^{T}\equiv\Phi_{\beta^{+}_{i}}$ Vandermonde vectors. The space spanned by $n$ $M$-dimensional Vandermonde vectors with different $\beta$ values is $n$-dimensional if $n<M$, and is  $M$-dimensional otherwise.

\begin{figure}
    \includegraphics[width=0.5\columnwidth]{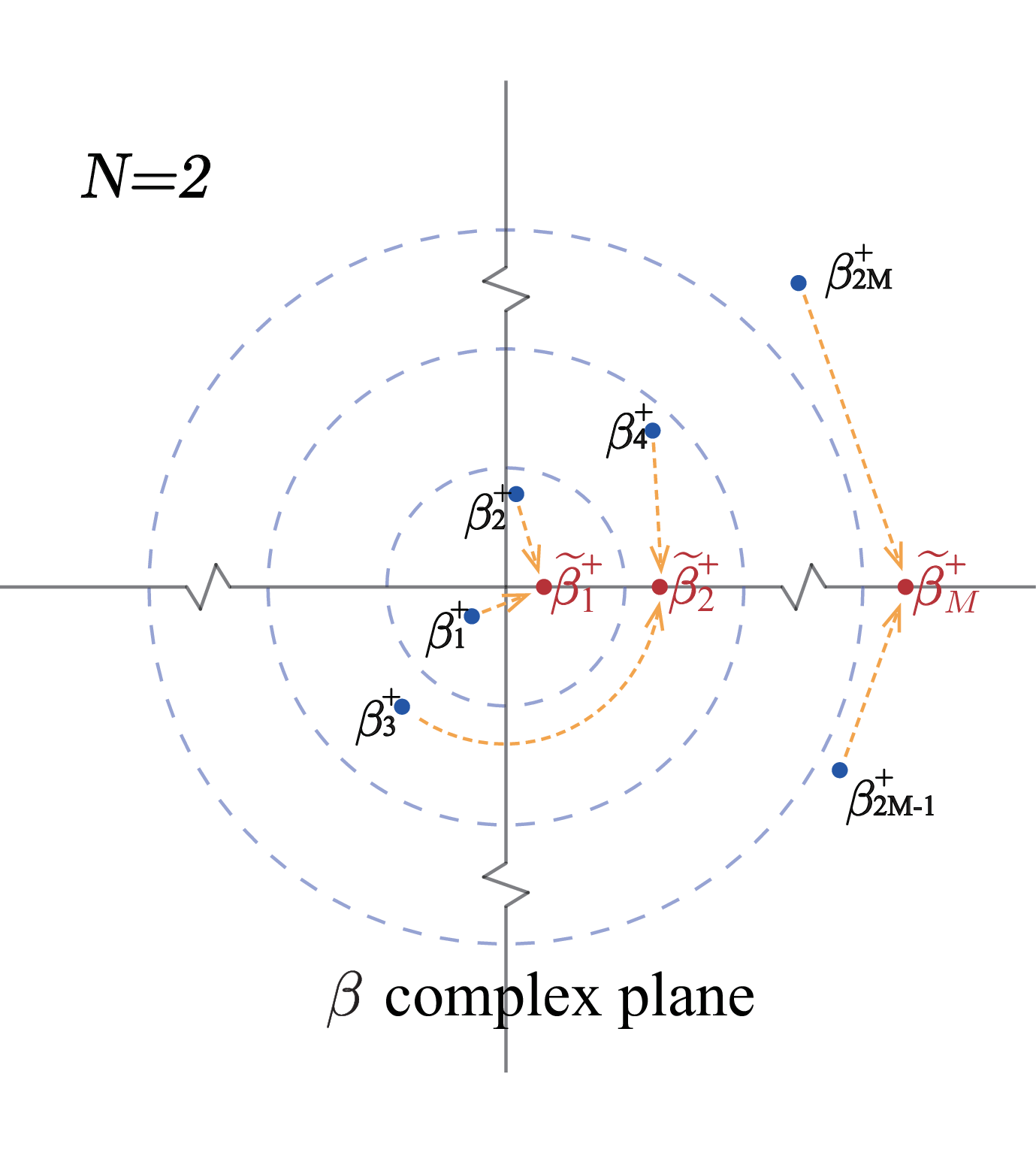}
    \caption{Continuous Variation of $R_{+}(\beta)$, where the $\beta$ roots of det$R_{+}(\beta)=0$ are varied continuously. The $NM$ roots are equally divided into $M$ groups with the $N$ roots in each group being finally merged into one repeated root, where $M=M_{1}+M_{2}$.}
    \label{sfig2}
\end{figure}

When $N\geq2$, each column vector $v_i$ has $N$ components and $R_+(\beta)$ is a $N\times N$ matrix. The $(kl)$ element of $R_+(\beta)$ is
a polynomial of $\beta$:
\begin{equation}
R_{kl}(\beta)=\sum_{m=-M_2}^{M_1}t^{kl}_{m}\beta^{-m},
\label{eq25}
\end{equation}
where $k$, $l$ =$1$, $2$,..., $N$. Design a set of continuous functions of $\Lambda$: $\beta^{+}_{i}(\Lambda)$, satisfying $\beta^{+}_{i}(0)=\beta^{+}_{i}$, and

\begin{equation}
 \begin{split}
&\beta^{+}_{1}(1)=\beta^{+}_{2}(1)=...=\beta^{+}_{N}(1)=\widetilde{\beta}^{+}_{1},\\
&\beta^{+}_{N+1}(1)=\beta^{+}_{N+2}(1)=...=\beta^{+}_{2N}(1)=\widetilde{\beta}^{+}_{2},\\
&\quad\quad\vdots\\
&\beta^{+}_{(M_{1}+M_{2}-1)N+1}(1)=\beta^{+}_{(M_{1}+M_{2}-1)N+2}(1)=...=\beta^{+}_{(M_{1}+M_{2})N}(1)=\widetilde{\beta}^{+}_{M_{1}+M_{2}}.
 \end{split}
 \label{eq26}
\end{equation}
Here $\widetilde{\beta}^{+}_{i}$ are the target $\beta$ with different absolute values satisfying: $\vert\beta^{+}_{(i-1)N+1}\vert\leq\vert\widetilde{\beta}^{+}_{i}\vert\leq\vert\beta^{+}_{iN}\vert$. So all $\beta^{+}_{i}(\Lambda)$ are expected to be divided into $M_{1}+M_{2}$ groups, and each group contains $N$ $\beta$ roots which would merge into one single $N$-fold root as $\Lambda$ is varied to be $1$. This is shown schematically in Fig.\ref{sfig2}. Now we treat all the coefficients $t^{kl}_{m}$ as unknown variables which can be viewed as functions of $\Lambda$: $t^{kl}_{m}(\Lambda)$. They can be found by solving the equations: det$R_{+}(\beta^{+}_{i},\Lambda)=0$, with $i=1$, $2$, ..., $N(M_{1}+M_{2})$ and imposing the initial conditions: $t^{kl}_{m}(0)=t^{kl}_{m}$. The number of the non-linear equations is $N(M_{1}+M_{2})$, while the number of the unknown variables is $N^{2}(M_{1}+M_{2}+1)$. So for any fixed value of $\Lambda$, we generally have multiple solutions of $t^{kl}_{m}(\Lambda)$, but for $\Lambda=0$, we have only one solution: $t^{kl}_{m}(0)=t^{kl}_{m}$. Thus we have found a solution of $R_{\pm}(\beta,\Lambda)$, which is a continuous function of $\Lambda$. When $\Lambda=1$, each $\widetilde{\beta}_{i}$ is a $N-$fold root of det$R_{+}(\beta^{+}_{i},\Lambda=1)=0$, indicating that its eigen-space should be $N$-dimensional and so $R_{+}(\beta^{+}_{i},\Lambda=1)\equiv0$. Therefore, as varying $\Lambda$ from $0$ to $1$, one would continuously transform $R_{+}(\beta,\Lambda)$ from $R_{+}(\beta)$ to $R_{+}(\beta,\Lambda=1)\propto\beta^{-NM_{1}}\prod_{1\leq i \leq M_{1}+M_{2}}(\beta-\widetilde{\beta}_{i}^{+})\times I_N$. When $\Lambda=1$, eigenvectors $v_{i}$ can be chosen to be $(v_{1},v_{2},...,v_{N(M_{1}+M_{2})})=(I_N, I_N, ..., I_N)$. It's easy to find that $\mathcal{M}_{N}^+(\Lambda=1)=\mathcal{\widetilde{M}}_{1}^+\otimes I_{N}$, where $\mathcal{\widetilde{M}}_{1}^+$ is the Vandermonde matrix of $(\widetilde{\beta}^{+}_{i})^{-1}$, and so  det$\mathcal{M}_{N}^+(\Lambda=1)=\mathcal{P}^{N}_{V}((\widetilde{\beta}_1^+)^{-1},(\widetilde{\beta}_2^+)^{-1},...,(\widetilde{\beta}^{+}_{M_{1}+M_{2}})^{-1})\neq0$

In the exactly similar way, from all the eigenvectors $u_{i}$ of $R_{-}$ one can construct the $N(M_1+M_2)\times N(M_1+M_2)$ matrix $\mathcal{M}_{N}^-$:
\begin{equation}
    \mathcal{M}_{N}^{-}=(\Phi^-_1,\Phi^-_2,...,\Phi^-_{N(M_1+M_2)}),
    \label{eq27}
\end{equation}
with the column vectors $\Phi^-_i=(u_i,(\beta_i^-)^{-1}u_i,...,(\beta_i^-)^{-M_1-M_2+1}u_i)^T=(1,(\beta_i^-)^{-1},...,(\beta_i^-)^{-M_1-M_2+1})^T\otimes u_i=\Phi_{\beta^{-}_{i}}\otimes u_i$. We also have: det$\mathcal{M}_{N}^-=\mathcal{P}^{N}_{V}((\widetilde{\beta}_1^-)^{-1},(\widetilde{\beta}_2^-)^{-1},...,(\widetilde{\beta}^{-}_{M_{1}+M_{2}})^{-1})\neq0$, where $\widetilde{\beta}_i^-$ are the target $\beta$ at $\Lambda=1$ in this case.

Under this situation, we discuss why the ranks of the matrices $\mathcal{M_L}$ and $\mathcal{M_R}$ in the main text have been maximized. We take $\mathcal{M_L}$ as an example to illustrate this fact. If $w=0$, then $p_{1}=p_{2}=p/2=NM_{1}$, $q_{1}=q_{2}=q/2=NM_{2}$. From $\mathcal{M}_{N}^+(\Lambda=1)$, choose the first $NM_{1}$ column vectors $\Phi^{+}_{i}=\Phi_{\beta^{+}_{i}}\otimes v_{i}$. For each vector we remove the last $NM_{2}$ components and replace $v_{i}$ by $\phi_{i}=(0,v_{i})^{T}$ to obtain $\Phi^{+}_{i,\mathcal{L}}$. These new vectors are comprised of $N$ sets of $M_{1}$-dimensional Vandermonde vectors with each set spanning an independent $M_{1}$-dimensional space. From $\mathcal{M}_{N}^-(\Lambda=1)$, also choose its first $NM_{1}$ column vectors $\Phi^{-}_{i}=\Phi_{\beta^{-}_{i}}\otimes u_{i}$. For each vector we also remove the last $NM_{2}$ components but replace $u_{i}$ by $\phi_{i}=(u_{i},0)^{T}$ to obtain $\Phi^{-}_{i,\mathcal{L}}$. These new vectors are also comprised of $N$ sets of $M_{1}$-dimensional Vandermonde vectors with each set spanning an independent $M_{1}$-dimensional space. Obviously, two spaces spanned by $\Phi^{+}_{i,\mathcal{L}}$ or $\Phi^{-}_{i,\mathcal{L}}$ are orthogonal. Since the matrix $\mathcal{M_L}=(\Phi_{1,\mathcal{L}}^+,...,\Phi_{NM_{1},\mathcal{L}}^+,\Phi_{1,\mathcal{L}}^-,...,\Phi_{NM_{1},\mathcal{L}}^-)$ is comprised of $\Phi^{+}_{i,\mathcal{L}}$ and $\Phi^{-}_{i,\mathcal{L       }}$, we have Rank$\mathcal{M_L}=$Rank$(\Phi_{1,\mathcal{L}}^+,...,\Phi_{NM_{1},\mathcal{L}}^+)$+Rank$(\Phi_{1,\mathcal{L}}^-,...,\Phi_{NM_{1},\mathcal{L}}^-)=NM_{1}+NM_{1}=p$. Thus, when $w=0$, Rank$\mathcal{M_L}$ is maximized to be $p$. Similarly, it can be shown that Rank$\mathcal{M_R}$ is maximized to be $q$.

When $w\neq0$, $p_{1}=NM_{1}+w$, $p_{2}=NM_{1}-w$, $q_{1}=NM_{2}-w$ and $q_{2}=NM_{2}+w$. We only consider the case of $0<w<N$ for simplicity. Let's make a slight modification on the continuous variation of $\Lambda$. We change the target $\beta$ roots slightly: only the $(M_{1}+1)$th equation of Eq.(12) is modified to be:
\begin{equation}
 \begin{split}
&\beta^{+}_{M_{1}N+1}(1)=\beta^{+}_{M_{1}N+2}(1)=...=\beta^{+}_{M_{1}N+w}(1)=\widetilde{\beta}^{+}_{M_{1}+1},\\
&\beta^{+}_{M_{1}N+w+1}(1)=\beta^{+}_{M_{1}N+w+2}(1)=...=\beta^{+}_{(M_{1}+1)N}(1)=\widetilde{\beta'}^{+}_{M_{1}+1},
 \end{split}
 \label{eq28}
\end{equation}
where $\vert\beta^{+}_{M_{1}N+1}\vert\leq\vert\widetilde{\beta}^{+}_{M_{1}+1}\vert\leq\vert\beta^{+}_{M_{1}N+w}\vert<\vert\beta^{+}_{M_{1}N+w+1}\vert\leq\vert\widetilde{\beta'}^{+}_{M_{1}+1}\vert\leq\vert\beta^{+}_{(M_{1}+1)N}\vert$. So here the $(M_{1}+1)$th group of $\beta^{+}_{i}(\Lambda)$ roots would merge into two repeated roots $\widetilde{\beta}^{+}_{M_{1}+1}$ and $\widetilde{\beta'}^{+}_{M_{1}+1}$, where the former is a $w$-fold root while the latter is a $(N-w)$-fold root. The eigenspace of the two roots formed by their corresponding eigenvectors $v_{i}$ is $w$- or $(N-w)$- dimensional, respectively. From $\mathcal{M}_{N}^+(\Lambda=1)$, choose the first $p_{1}$ column vectors $\Phi^{+}_{i}=\Phi_{\beta^{+}_{i}}\otimes v_{i}$. For each vector we remove the last $NM_{2}$ components and replace $v_{i}$ by $\phi_{i}=(0,v_{i})^{T}$, so we obtain $\Phi^{+}_{i,\mathcal{L}}$. These new vectors have already contained $N$ sets of $M_{1}$-dimensional Vandermonde vectors with each set spanning an independent $M_{1}$-dimensional space. So Rank$(\Phi_{1,\mathcal{L}}^+,...,\Phi_{p_1,\mathcal{L}}^+)$ has been maximized to be $NM_{1}=p/2$. From $\mathcal{M}_{N}^-(\Lambda=1)$, also choose its first $p_{2}$ column vectors $\Phi^{-}_{i}=\Phi_{\beta^{-}_{i}}\otimes u_{i}$. For each vector we also remove the last $NM_{2}$ components but replace $u_{i}$ by $\phi_{i}=(u_{i},0)^{T}$, so we obtain $\Phi^{-}_{i,\mathcal{L}}$. If proper representation of $v_{i}$ is chosen, these new vectors are found to be comprised of $N-w$ sets of $M_{1}$-dimensional Vandermonde vectors with each set spanning an independent $M_{1}$-dimensional space, together with $w$ sets of $(M_{1}-1)$-dimensional Vandermonde vectors with each set spanning an independent $(M_{1}-1)$-dimensional space. So Rank$(\Phi_{1,\mathcal{L}}^-,...,\Phi_{p_2,\mathcal{L}}^-)=NM_{1}-w=p/2-w$. Therefore, we have Rank$\mathcal{M_L}=$Rank$(\Phi_{1,\mathcal{L}}^+,...,\Phi_{p_1,\mathcal{L}}^+)$+Rank$(\Phi_{1,\mathcal{L}}^-,...,\Phi_{p_2,\mathcal{L}}^-)=p/2+(p/2-w)=p-w$, indicating that it has been maximized. Thus, when $w>0$, Rank$\mathcal{M_L}$ is maximized to be $p-w$. Similarly, it can be shown that Rank$\mathcal{M_R}$ is simultaneously maximized to be $q-w$.

So we have proved that for a state of the open system denoted by $w$ in the main text, \textit{by continuously varying $R_{\pm}(\beta)$ without closing the gap, the ranks of $\mathcal{M}_L$ and $\mathcal{M}_R$ in the main text can be maximized to be $p-\vert w\vert$ and $q-\vert w\vert$ respectively.}

\quad

\quad

\centerline{\textbf{\Rmnum{3}.A Theorem on $\pm1$ change at topological transitions of the Winding Number Defined on One of the SubGBZs}}

\quad

In this section we prove that when the non-hermitian system is sufficiently near the topological transition point, if one replaces the closed loop $\mathcal{L_{\beta}}$ in the integral of the topological invariant $\nu$ defined in the main text, by  one of the subGBZs, the change of the new winding number $\nu^{'}$ at the transition point would be $\Delta\nu^{'}=\pm1$. The subGBZ chosen corresponds to the branch of the OBS closest to the $E=0$ point.

\begin{figure}
    \includegraphics[width=0.6\columnwidth]{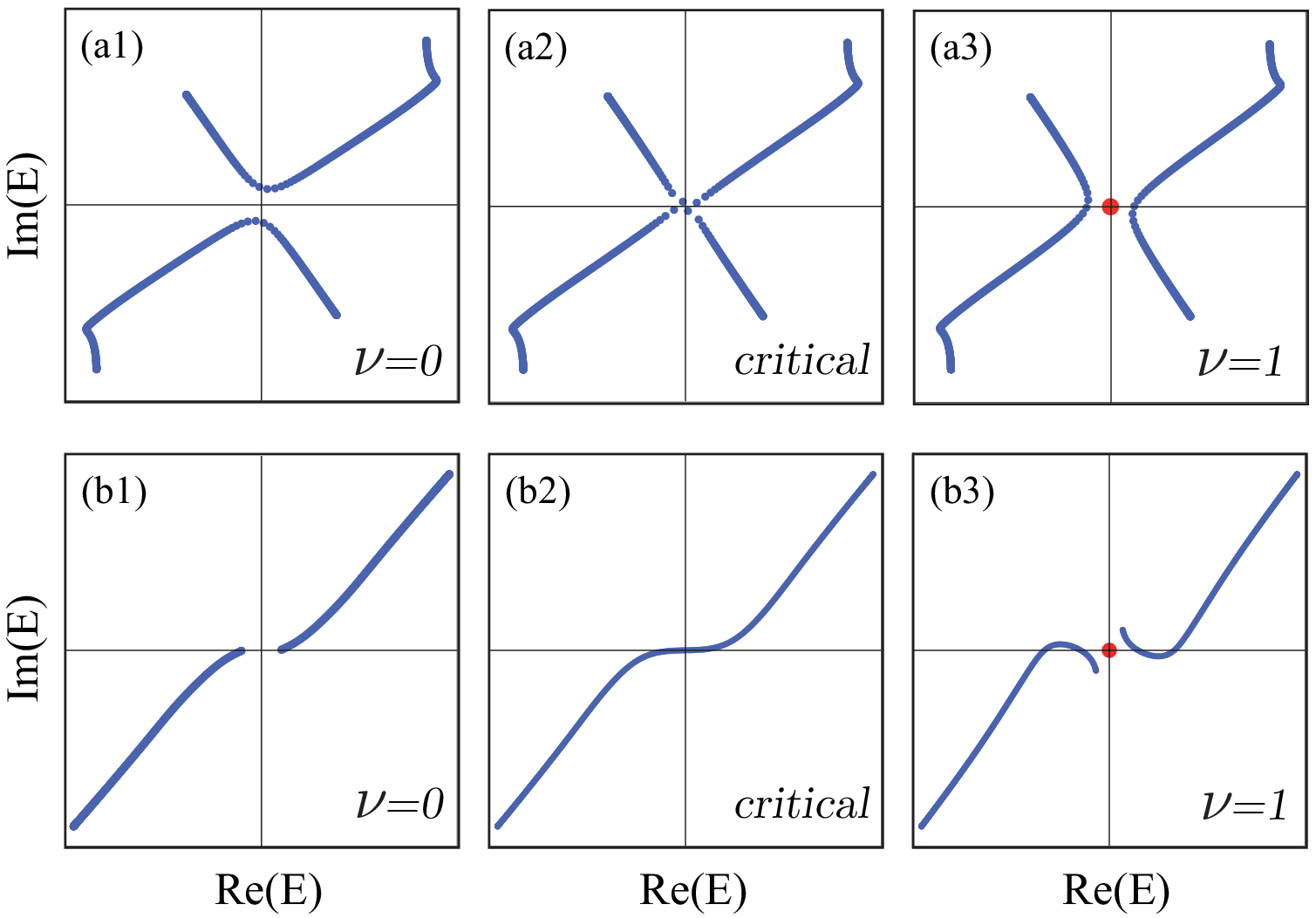}
    \caption{Two typical topological transition processes under parameter change in a chiral symmetric non-hermitian system under OBCs, where gap closes and then opens. Each pair of arcs in each figure represents the corresponding pair of the OBSs on $E$ plane closest to $E=0$. The isolated solid circle at $E=0$ in (a3) and (b3) denotes the topological protected zero modes with $\nu$ denoting the corresponding topological invariant. (a1)-(a3): innerpoint touching process. (b1)-(b3): endpoint touching process.}
    \label{sfig3}
\end{figure}

When the system under OBCs is near the topological phase transition, the system must be nearly gapless and then one part of its OBS must be approaching the $E=0$ point. Under parameter change, the transition process can be summarized into two scenarios, as shown in Fig.\ref{sfig3}, where one is the innerpoint connection and the other is the endpoint connection. At the critical point where the system is gapless, as shown in Fig.\ref{sfig3} (a2) and (b2), the state of the isolated point $E=0$ belongs to one pair of the open arcs of the OBS and its $\beta_p(0)$ and $\beta_{p+1}(0)$ would simultaneously fall on the corresponding subGBZ. Now consider the situation when the system is sufficiently close to the transition point.
The subGBZ as a closed loop, is composed of the $\beta_{p}$ and $\beta_{p+1}$ points of the energy branch of the OBS closest to $E=0$. Choose a special state $E_{0}$ on the branch which is nearest to $E=0$. If the system is sufficiently close to the transition point, then $E_{0}$ would be approaching $E=0$ so that its $\beta_{j}(E_{0})$ is nearly equal to $\beta_{j}(0)$.

\begin{figure}
    \includegraphics[width=0.9\columnwidth]{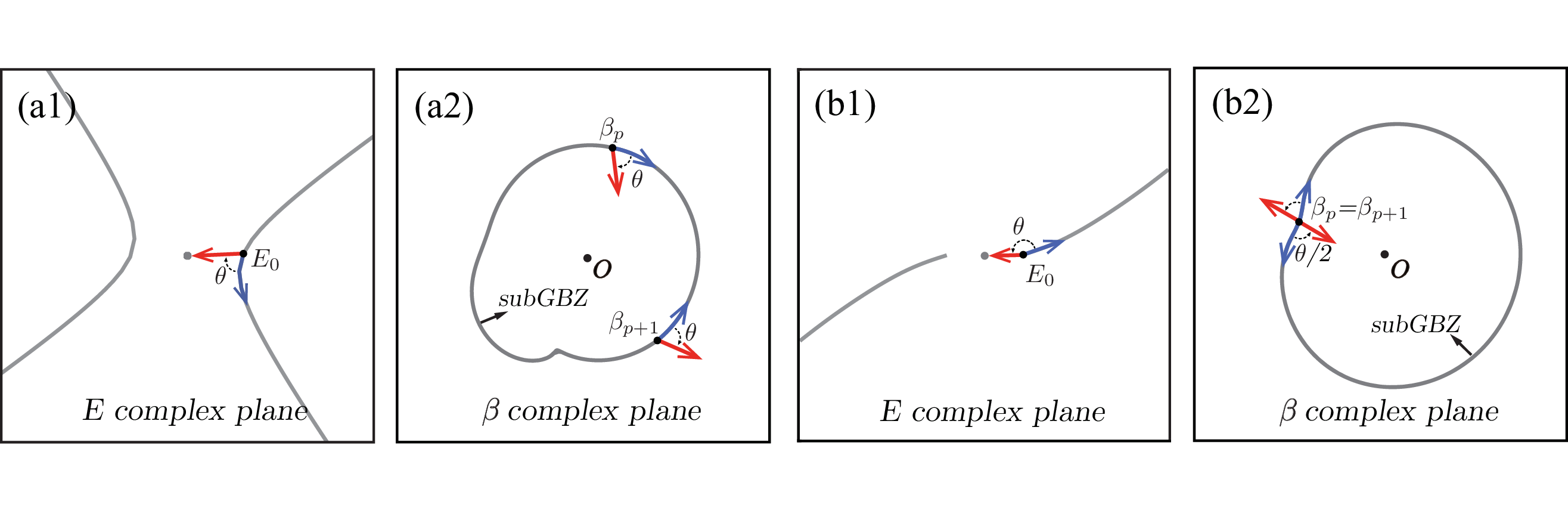}
    \caption{Two deviation paths in $E$ plane and the corresponding paths of their $\beta_p$ and $\beta_{p+1}$ in $\beta$ plane for the two cases when the system is sufficiently near the critical point, as shown in Fig.3. Here, $E_0$ is the point on the OBS which is chosen to be the nearest one to $E=0$. One path on $E$ plane is the straight line segment connecting $E_0$ to $E=0$, the other path starting from $E_0$ is a small arc along the same OBS. }
    \label{sfig4}
\end{figure}

Let's first consider the case where the innerpoints $\pm E_{0}$ are going to be touching with each other, as shown in Fig.3(a1)-(a2). Expand the polynomial $f(E,\beta)\equiv$det$(H(\beta)-E)=0$ near $E_0$ and its $\beta_{j}(E_{0})$. We have
\begin{equation}
\frac{\partial f}{\partial E}\bigg|_{(E_{0},\beta_{j}(E_{0}))}\Delta E+ \frac{\partial f}{\partial \beta}\bigg|_{(E_{0},\beta_{j}(E_{0}))}\Delta\beta=0.
\label{eq31}
\end{equation}
So the small deviations $\Delta E$ and $\Delta\beta$ are connected with each other. For any two kinds of deviations ($\Delta E,\Delta\beta$), and ($\Delta E^{'},\Delta\beta^{'}$), one must have
\begin{equation}
    {\Delta E^{'}}/{\Delta E}={\Delta \beta^{'}}/{\Delta \beta}.
    \label{eq32}
\end{equation}
This is actually the conformal property of analytic functions in complex analysis. Now take $\Delta E$ as the deviation along the OBS, and $\Delta E^{'}$ as the deviation along the line segment, as shown in Fig.\ref{sfig4} (a1). If $j$ is chosen to be $p$ or $p+1$, we would have two deviations $\Delta\beta$, starting from $\beta_{p}$ or $\beta_{p+1}$ respectively, as shown in Fig.\ref{sfig4} (a2). What's important is that $\Delta\beta$ at $\beta_{p}$ and $\Delta\beta$ at $\beta_{p+1}$ are pointing oppositely along the subGBZ\cite{PhysRevB.105.045422}, namely, one is clockwise while the other is anticlockwise. Therefore, according to the conformal property Eq.(\ref{eq32}), the deviation $\Delta\beta^{'}$ at $\beta_{p}$($\beta_{p+1}$) would be pointing towards the interior(exterior) of the subGBZ. So among two points $\beta_{p}(0)$ and $\beta_{p+1}(0)$, the subGBZ only encloses $\beta_{p}(0)$.
For the case of the endpoints $\pm E_0$ touching with each other, as shown in Fig.\ref{sfig4} (b1)-(b2), since $\beta_{p}(E_{0})=\beta_{p+1}(E_{0})$, one has $\partial f/\partial\beta\mid_{(E_{0},\beta_{p}(E_{0}))}=\partial f/\partial\beta\mid_{(E_{0},\beta_{p+1}(E_{0}))}=0$. So the expansion of the polynomial $f(E,\beta)$ near $E_0$ and its $\beta_{p}(E_{0})$ must be made in the second order of $\beta$ and we obtain:
\begin{equation}
    \begin{split}
        &\frac{\partial f}{\partial E}\bigg|_{(E_{0},\beta_{p}(E_{0}))}\Delta E+ \frac{1}{2}\frac{\partial^2 f}{\partial \beta^2}\bigg|_{(E_{0},\beta_{p}(E_{0}))}\Delta \beta^2=0.\label{eq33}
    \end{split}
\end{equation}
So similarly for two kinds of deviations, we must have:
\begin{equation}
       {\Delta E^{'}}/{\Delta E}=({\Delta \beta^{'}}/{\Delta \beta})^2,
       \label{eq34}
\end{equation}
where the two kinds of deviations $\Delta E$ and $\Delta E^{'}$ are taken similarly as before, as shown in Fig.\ref{sfig4} (b1). The two corresponding deviations of $\Delta\beta^{'}$ is found to be pointing oppositely. Thus for $\beta_{p}(0)$ and $\beta_{p+1}(0)$ points, the subGBZ only encircles $\beta_{p}(0)$.

Therefore for both cases, we have proved that among $\beta_{p}(0)$ and $\beta_{p+1}(0)$, the subGBZ as a closed loop only encircles $\beta_{p}(0)$, with $\beta_{p+1}(0)$ being kept outside. Consider a topological transition process with $\Delta\nu=\pm1$. Before or after the transition, the system would always be gapful, but is characterized by different topological invariant $\nu$. Let's denote $\beta^{i}_{p}(0)$($\beta^{f}_{p}(0)$) and $\beta^{i}_{p+1}(0)$($\beta^{f}_{p+1}(0)$) for the system just before(after) the phase transition. According to our theorem, at the transition point, one of the $\beta_{p}(0)$ and $\beta_{p+1}(0)$ points is from det$R_{+}=0$, while the other is from det$R_{-}=0$, and they would exchange to induce the topological transition. So during the phase transition, $\beta^{i}_{p}(0)$ would continuously change to be $\beta^{f}_{p+1}(0)$, while $\beta^{i}_{p+1}(0)$ would continuously change to be $\beta^{f}_{p}(0)$. Then the change of the phase winding along the subGBZ for det$R_{+}$ would be minus that for det$R_{-}$. This indicates that the change of the new winding number $\nu^{'}$ defined on the subGBZ would be $\pm1$. So we complete the proof of the statement at the beginning of this section. We remark that for a two-band system, its GBZ naturally becomes the subGBZ discussed here.

\quad

\quad

\centerline{\textbf{\Rmnum{4}.Explicit Model Example Exhibiting Relationship between the Invariant $\nu$ and the Edge States}}

\quad

In this section, by studying an explicit model in detail, we will follow the way in the theorem proof in the main text, step by step, to find out the edge states for the model system under OBCs. On the other hand, by direct diagonalization of the Hamiltionian for the corresponding open system, we obtain the numerical results on the edge states of $E=0$. In the following, we will compare these numerical results with our analytical calculations, to illustrate our conclusions on the topologically protected edge states(TPESs) given in the main text.

\begin{figure}
    \centering
    \includegraphics[width=0.6 \columnwidth]{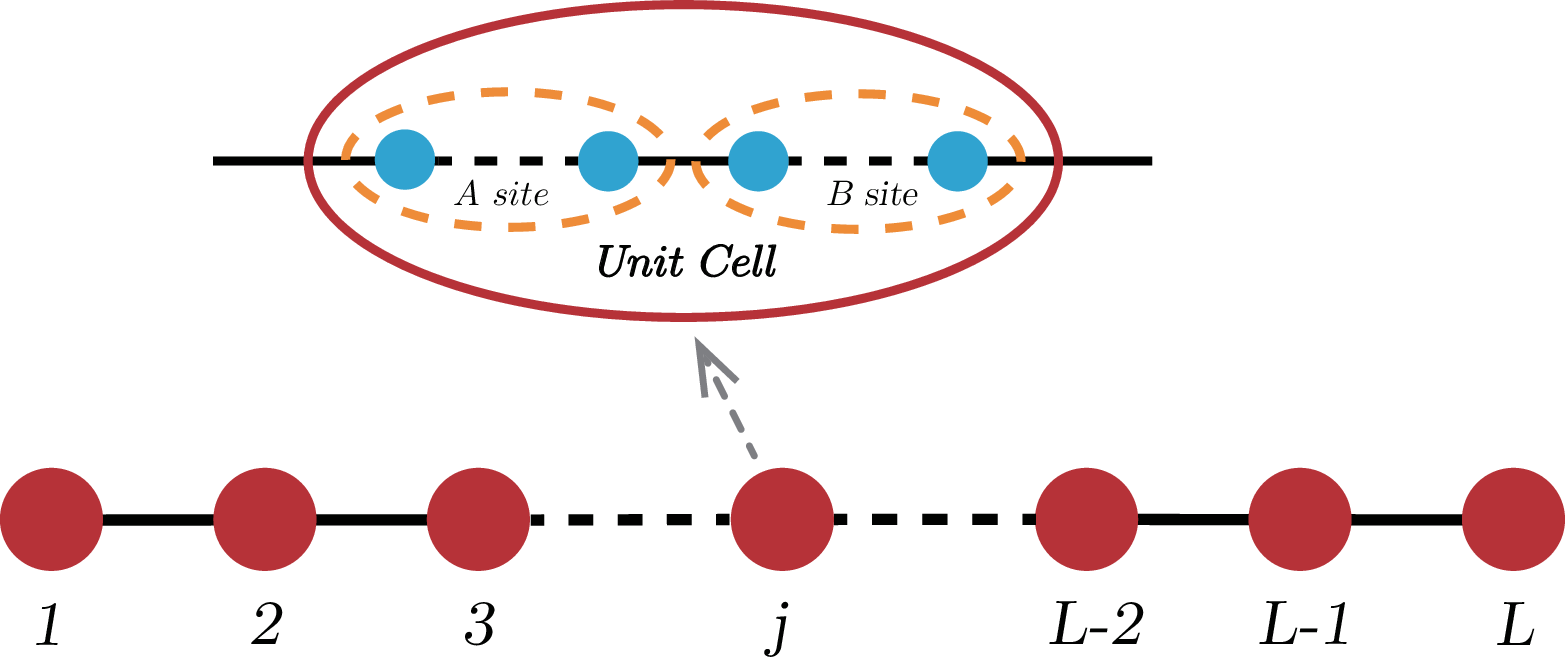}
    \caption{Open chain and its unit cells which can be divided into two parts: sites belonging to sublattice A and those belonging to sublattice B.}
    \label{sfig5}
\end{figure}

When a $2N$-band non-hermitian system with chiral symmetry takes the form Eq.(3) in the main text, it is convenient to introduce the two sublattices A and B, where the first(last) $N$ internal degrees of freedom belong to sublattice A(B), as shown in Fig.\ref{sfig5}. We choose the model in the main text, which is a four-band model with only the nearest-neighbor hoppings, implying $N=2$, $M_{1}=M_{2}=1$ and $p=q=4$.\\
\indent
First, the parameter $\lambda$ is chosen to be $\lambda=0.1$, corresponding to a topologically nontrivial state with $\nu=-1$. By solving det$H(\beta)=0$, we get eight $\beta$ roots of $E=0$, which are given by:
\begin{equation}
    \begin{split}
        \begin{array}{cccc}
        \beta_1=\beta_1^-=-0.195, &\quad\beta_2=\beta_1^+=-0.401, &\quad\beta_3=\beta_2^-=-0.495, &\beta_4=\beta_3^-=0.811,\\
        \beta_5=\beta_2^+=-1.149, &\beta_6=\beta_3^+=2.014, &\quad\beta_7=\beta_4^-=-2.442, &\quad\beta_8=\beta_4^+=-5.634,
        \end{array}
    \end{split}
    \label{eq41}
\end{equation}
and accordingly their corresponding eigenmodes are:
\begin{equation}
    \begin{split}
        \begin{array}{cccc}
    \phi_1=\phi_1^-=\left(
            \begin{array}{c}
               -0.569 \\
               -0.822\\
               0\\
               0 \\
            \end{array}
            \right),\quad
    &\phi_2=\phi_1^+=\left(
            \begin{array}{c}
               0 \\
               0\\
               -0.664\\
               -0.748 \\
            \end{array}
            \right),\quad
    &\phi_3=\phi_2^-=\left(
            \begin{array}{c}
               -0.084 \\
               -0.997\\
               0\\
               0 \\
            \end{array}
            \right),\quad
    &\phi_4=\phi_3^-=\left(
            \begin{array}{c}
               -0.464 \\
               0.886\\
               0\\
               0 \\
            \end{array}
            \right),\\
    \phi_5=\phi_2^+=\left(
            \begin{array}{c}
                0 \\
                0\\
                0\\
                1.000 \\
            \end{array}
            \right),\quad\quad
    &\quad\phi_6=\phi_3^+=\left(
            \begin{array}{c}
                0 \\
                0\\
                -0.409\\
                0.912 \\
            \end{array}
            \right),\quad\quad
    &\phi_7=\phi_4^-=\left(
            \begin{array}{c}
            -0.714 \\
            -0.700\\
            0\\
            0 \\
        \end{array}
        \right),\quad
    &\phi_8=\phi_4^+=\left(
        \begin{array}{c}
            0 \\
            0\\
            -0.648\\
            -0.761 \\
        \end{array}
        \right).\quad
    \end{array}
    \end{split}
    \label{eq42}
\end{equation}
Here the superscript $\pm$ indicates that the corresponding $\beta$ roots come from $R_{\pm}$. So the matrices $\mathcal{M_L}=(\phi_1^+,\phi_1^-,\phi_2^-,\phi_3^-)$ and $\mathcal{M_R}=(\beta_2^+\phi_2^+,\beta_3^+\phi_3^+,\beta_4^+\phi_4^+,\beta_4^-\phi_4^-)$ can be explicitly expressed as
\begin{equation}
    \begin{split}
        \mathcal{M_L}=\left(
\begin{array}{cccc}
 0 & -0.569 &
   -0.084 & -0.464
   \\
 0 & -0.822 & 0.996 & 0.886 \\
 -0.664 & 0 & 0 & 0 \\
 -0.748 & 0 & 0 & 0 \\
\end{array}
\right),\quad
\mathcal{M_R}=\left(
    \begin{array}{cccc}
     0& 0& 0& 1.744 \\
     0& 0& 0&
       1.710 \\
     0 & -0.824
       & 3.649 & 0 \\
     -1.149 & 1.840 & 4.292 & 0 \\
    \end{array}
    \right).
    \end{split}
    \label{eq43}
\end{equation}
It's easy to check that Rank$\mathcal{M_L}=$Rank$\mathcal{M_R}=3$, namely, they have been maximized so either $\mathcal{M_L}\vert \mathcal{L}\rangle=0$ or $\mathcal{M_R}\vert \mathcal{R}\rangle=0$ has one independent solution, which can be expressed as $\vert \mathcal{L}\rangle =(c_1^+,c_1^-,c_2^-,c_3^-)^T$ and $\vert \mathcal{R}\rangle =(c_2^+,c_3^+,c_4^+,c_4^-)^T$. The solutions are found to be: $\vert \mathcal{L}\rangle =(0, 0.335, 0.765, -0.550)^T$ and  $\vert \mathcal{R}\rangle =(0.922, 0.377, 0.0852, 0)^T$. Notice that in these expressions the decimal numbers shown here are all retained to a few decimal places, but in the actual calculations they have all been kept in high precision to ensure the accuracy of the calculating results.

\begin{figure}
    \centering
    \includegraphics[width=0.9\columnwidth]{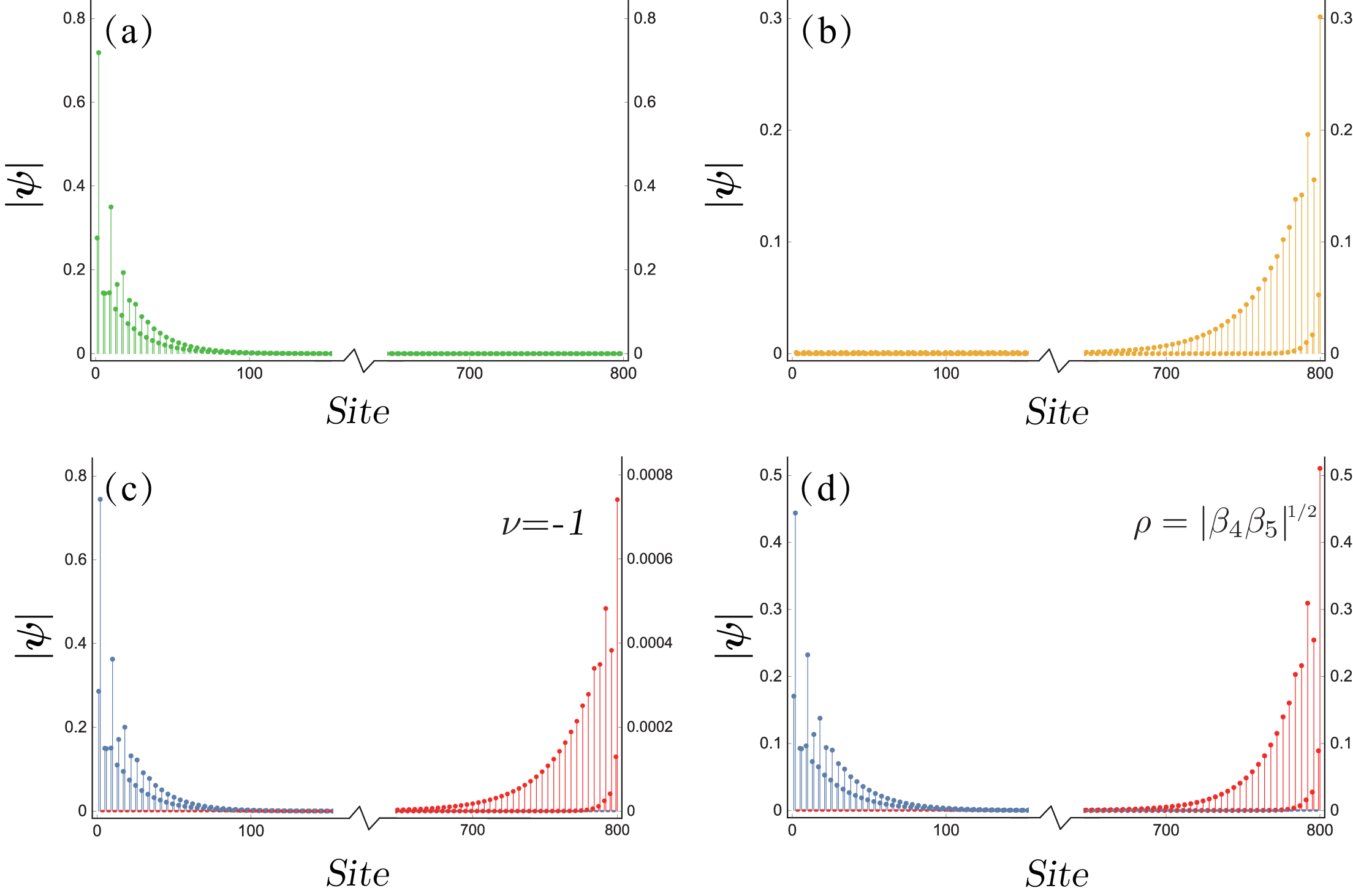}
    \caption{Space distributions of the edge states with $\nu=-1$ and $\lambda=0.1$. (a) Edge state $\psi_{\mathcal{L}}$. (b) Edge state $\psi_{\mathcal{R}}$. (c) Numerical matrix diagonalization result of the zero-energy modes, and (d) the corresponding result after a similarity transformation with $\rho=\vert\beta_4\beta_5\vert^{1/2}$, where data on sublattice A(B) are colored in blue(red). We have 800 lattice sites since here $L=200$ and $N=2$.}
    \label{sfig6}
\end{figure}

From the two solutions one can see that $\vert \mathcal{L}\rangle$ only contains the coefficients of $\phi^-$, while $\vert \mathcal{R}\rangle$ only contains those of $\phi^+$. This indicates that $\vert \mathcal{L}\rangle$ is a left-localized state residing on sublattice A, while $\vert \mathcal{R}\rangle$ is a right-localized one residing on sublattice B. By using these coefficients, the corresponding real-space normalized wave function $\psi_{\mathcal{L}}$ and $\psi_{\mathcal{R}}$ can be written as:
\begin{equation}
    \begin{split}
       &\psi_{\mathcal{L}}(j)\propto \sum_{j=1}^L c_1^-(\beta_1^{-})^j\phi_1^-+c_2^-(\beta_2^{-})^j\phi_2^-+c_3^-(\beta_3^{-})^j\phi_3^-,\\
       &\psi_{\mathcal{R}}(j)\varpropto \sum_{j=1}^L c_2^+(\beta_2^{+})^{j-L}\phi_2^++c_3^+(\beta_3^{+})^{j-L}\phi_3^++c_4^+(\beta_4^{+})^{j-L}\phi_4^+.
    \end{split}
    \label{eq44}
\end{equation}
The space distributions of these two wave functions are shown in Fig.\ref{sfig6} (a)-(b). For a finite system under OBCs, we have two final real-space wave functions $\psi_{\pm}$ which are the linear combinations of $\psi_{\mathcal{L}}$ and $\psi_{\mathcal{R}}$:
\begin{equation}
    \psi_{\pm}=\mathcal{A}\psi_{\mathcal{L}}\pm\mathcal{B}\psi_{\mathcal{R}},
    \label{eq45}
\end{equation}
where $\psi_{+}$ and $\psi_{-}$ are related by chiral symmetry: $\mathcal{S}\psi_{\pm}=-\mathcal{A}\psi_{\mathcal{L}}\pm\mathcal{B}\psi_{\mathcal{R}}=-\psi_{\mp}$, with $\mathcal{S}$ the chiral operator. The coefficients $\mathcal{A}$ and $\mathcal{B}$ are determined by the matrices $\mathcal{M_{RL}}$ and $\mathcal{M_{LR}}$, which take the following forms:
\begin{equation}
    \begin{split}
&\mathcal{M_{RL}}=((\beta_1^+)^{L+1}\phi_1^+,(\beta_1^-)^{L+1}\phi_1^-,(\beta_2^-)^{L+1}\phi_2^-,(\beta_3^-)^{L+1}\phi_3^-),\\
&\mathcal{M_{LR}}=((\beta_2^+)^{-L}\phi_2^+,(\beta_3^+)^{-L}\phi_3^+,(\beta_4^+)^{-L}\phi_4^+,(\beta_4^-)^{-L}\phi_4^-).
    \end{split}
    \label{eq46}
\end{equation}
Substitute $\vert \mathcal{L}\rangle =(0, 0.335, 0.765, -0.550)^T$ and  $\vert \mathcal{R}\rangle =(0.922, 0.377, 0.0852, 0)^T$ into the equation
\begin{equation}
    \frac{\mathcal{B}}{\mathcal{A}}=\frac{\sqrt{\langle \mathcal{R}\vert\mathcal{M_{RL}\vert L \rangle}}}{\sqrt{\langle \mathcal{L}\vert\mathcal{M_{LR}\vert R \rangle}}},
    \label{eq47}
\end{equation}
one can find $\vert\mathcal{B}/\mathcal{A}\vert\approx \vert\beta^{-}_3\beta^{+}_2\vert^{L/2}=\vert\beta_4\beta_5\vert^{L/2}$. Here since $\vert\beta_4\beta_5\vert<1$, when $L$ is sufficiently large, it will be $\vert\mathcal{B}\vert\ll\vert\mathcal{A}\vert$, indicating the contribution of $\psi_{\mathcal{L}}$ to $\psi_{\pm}$ is much greater than that of $\psi_{\mathcal{R}}$. So for this parameter, it looks like that the two TPESs are all localized at the left boundary, and when $L\rightarrow\infty$, $\vert\mathcal{B}\vert$ will be approaching zero so that $\psi_{+}$ and $\psi_{-}$ linked by chiral symmetry would coalesce into one single state $\psi_{\mathcal{L}}$. This is actually the origin of the defectiveness of the TPESs. In Fig.\ref{sfig6} (c) we show the numerical matrix diagonalization result of the TPESs. The two edge states are nearly overlapping, and are found to be localized at left boundary while at right boundary the wave function only has a very small weight. This is consistent with our analysis. If we perform a similarity transformation on this model to make $\vert\beta_4\beta_5/\rho^2\vert=1$, i.e., $\rho=\vert\beta_4\beta_5\vert^{1/2}$, then this model described by $H(\beta)$ would be transformed to one described by $H(\rho\beta)$ whose wave function $\widetilde{\psi}(j)=\rho^{-j}\psi(j)$ and so its ratio $\vert\mathcal{B}/\mathcal{A}\vert\approx 1^L=1$. In our analysis the two TPESs in this situation would be localized at both ends of the open chain, consistent with the numerical results as shown in Fig.\ref{sfig6} (d).
Meanwhile, notice that a negative $\nu=-1$ means $\vert \mathcal{L}\rangle$ only has $\phi^-$ components, while $\vert \mathcal{R}\rangle$ only has $\phi^+$ ones, indicating that $\vert \mathcal{L}\rangle$($\vert \mathcal{R}\rangle$) is a left(right)-localized A(B)-sublattice edge state. However, for a positive $\nu=1$, the $\vert \mathcal{L}\rangle$($\vert \mathcal{R}\rangle$) state would be a left(right)-localized B(A)-sublattice edge state, as mentioned in the main text. This is because that when $\nu=+1$, all nonzero components in $\vert \mathcal{L}\rangle$($\vert \mathcal{R}\rangle$) are $\phi^+$($\phi^-$) instead of $\phi^-$($\phi^+$).
\begin{figure}
    \centering
    \includegraphics[width=0.9\columnwidth]{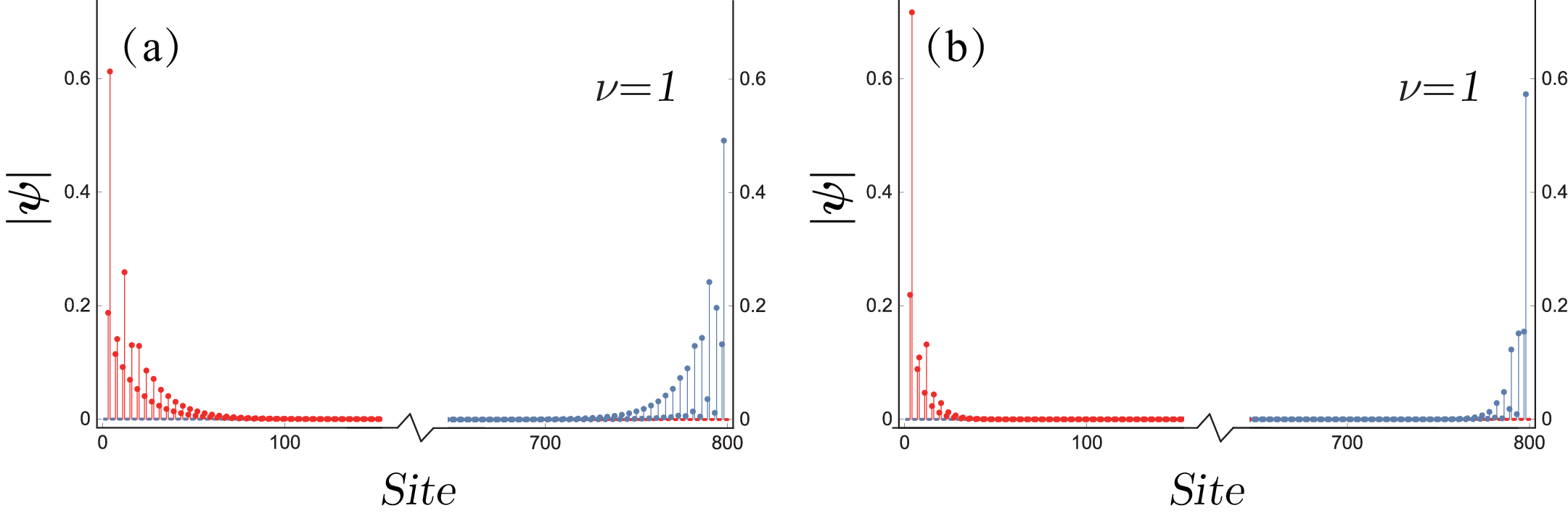}
    \caption{Space distributions of the zero-energy modes obtained by the numerical matrix diagonalization, with $\nu=1$, where data on sublattice A(B) are colored in blue(red). (a) $\lambda=-0.6$. (b) $\lambda=-0.9$. Here $L=200$. }
    \label{sfig7}
\end{figure}

\begin{figure}
    \centering
    \includegraphics[width=0.9\columnwidth]{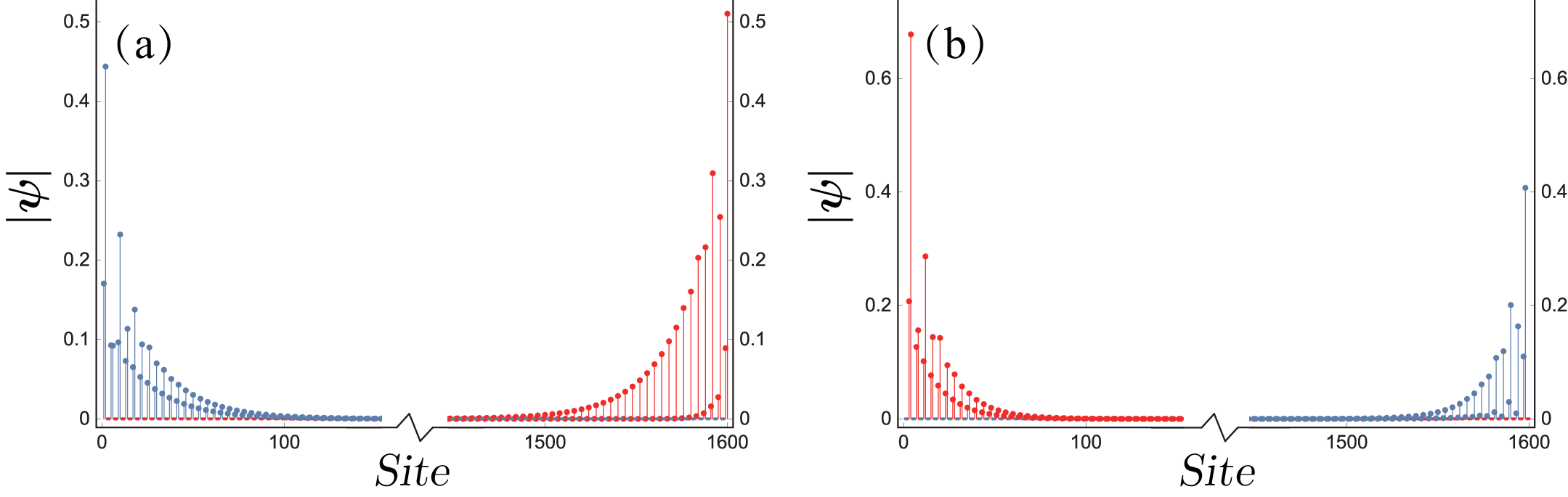}
    \caption{(a) The same as Fig.6 (d), and (b) the same as Fig.7 (a), except that $L=400$.}
    \label{sfig8}
\end{figure}

In Fig.\ref{sfig7}, we show the case of $\nu=1$ with $\lambda=-0.6$ and $\lambda=-0.9$ respectively. The eight $\beta$ values for the two parameters are ordered as usual as follows:
\begin{equation}
    \begin{split}
&\vert\beta_{1}\vert=\vert\beta^{+}_{1}\vert<\vert\beta_{2}\vert=\vert\beta^{+}_{2}\vert<\vert\beta_{3}\vert=\vert\beta^{-}_{1}\vert<\vert\beta_{4}\vert=\vert\beta^{+}_{3}\vert<1<\vert\beta_{5}\vert=\vert\beta^{-}_{2}\vert<\vert\beta_{6}\vert=\vert\beta^{+}_{4}\vert<\vert\beta_{7}\vert=\vert\beta^{-}_{3}\vert<\vert\beta_{8}\vert=\vert\beta^{-}_{4}\vert,\\
&\vert\beta_{1}\vert=\vert\beta^{+}_{1}\vert<\vert\beta_{2}\vert=\vert\beta^{+}_{2}\vert<\vert\beta_{3}\vert=\vert\beta^{+}_{3}\vert<\vert\beta_{4}\vert=\vert\beta^{-}_{1}\vert<1<\vert\beta_{5}\vert=\vert\beta^{+}_{4}\vert<\vert\beta_{6}\vert=\vert\beta^{-}_{2}\vert<\vert\beta_{7}\vert=\vert\beta^{-}_{3}\vert<\vert\beta_{8}\vert=\vert\beta^{-}_{4}\vert,
    \end{split}
    \label{eq48}
\end{equation}
where the former corresponds to $\lambda=-0.6$ while the latter corresponds to $\lambda=-0.9$. When $\lambda=-0.6$, $\vert\mathcal{B}/\mathcal{A}\vert\approx \vert\beta_4\beta_5\vert^{L/2}=\vert\beta^{+}_3\beta^{-}_2\vert^{L/2}$ and when $\lambda=-0.9$, $\vert\mathcal{B}/\mathcal{A}\vert\approx \vert\beta_3\beta_6\vert^{L/2}=\vert\beta^{+}_3\beta^{-}_2\vert^{L/2}$. In both cases, $\vert\mathcal{B}/\mathcal{A}\vert\approx 0.997^{L}$. In our numerical calculation, since $L=200$, $\vert\mathcal{B}/\mathcal{A}\vert\approx 0.74$, which makes $\vert\mathcal{A}\vert$ slightly larger than $\vert\mathcal{B}\vert$. So each TPES is localized at both boundaries of the open chain. But when $L\rightarrow\infty$, only the left-localized components would be remaining. By setting $L=400$ we redo the numerical calculation of the edge states in Fig.\ref{sfig6} (d) and Fig.\ref{sfig7} (a), and show the corresponding results in Fig.\ref{sfig8}. In Fig.\ref{sfig8} (a), the space distribution is nearly unchanged since $\vert\mathcal{B}/\mathcal{A}\vert\propto1$ is also kept unchanged while in Fig.\ref{sfig8} (b), the left-localized components are even more dominant than the right-localized ones since the ratio $\vert\mathcal{B}/\mathcal{A}\vert$ in this case becomes smaller. These results show that the predicted ratio between the coefficients $\mathcal{A}$ and $\mathcal{B}$ is quite reasonable.

\begin{figure}
    \centering
    \includegraphics[width=0.9\columnwidth]{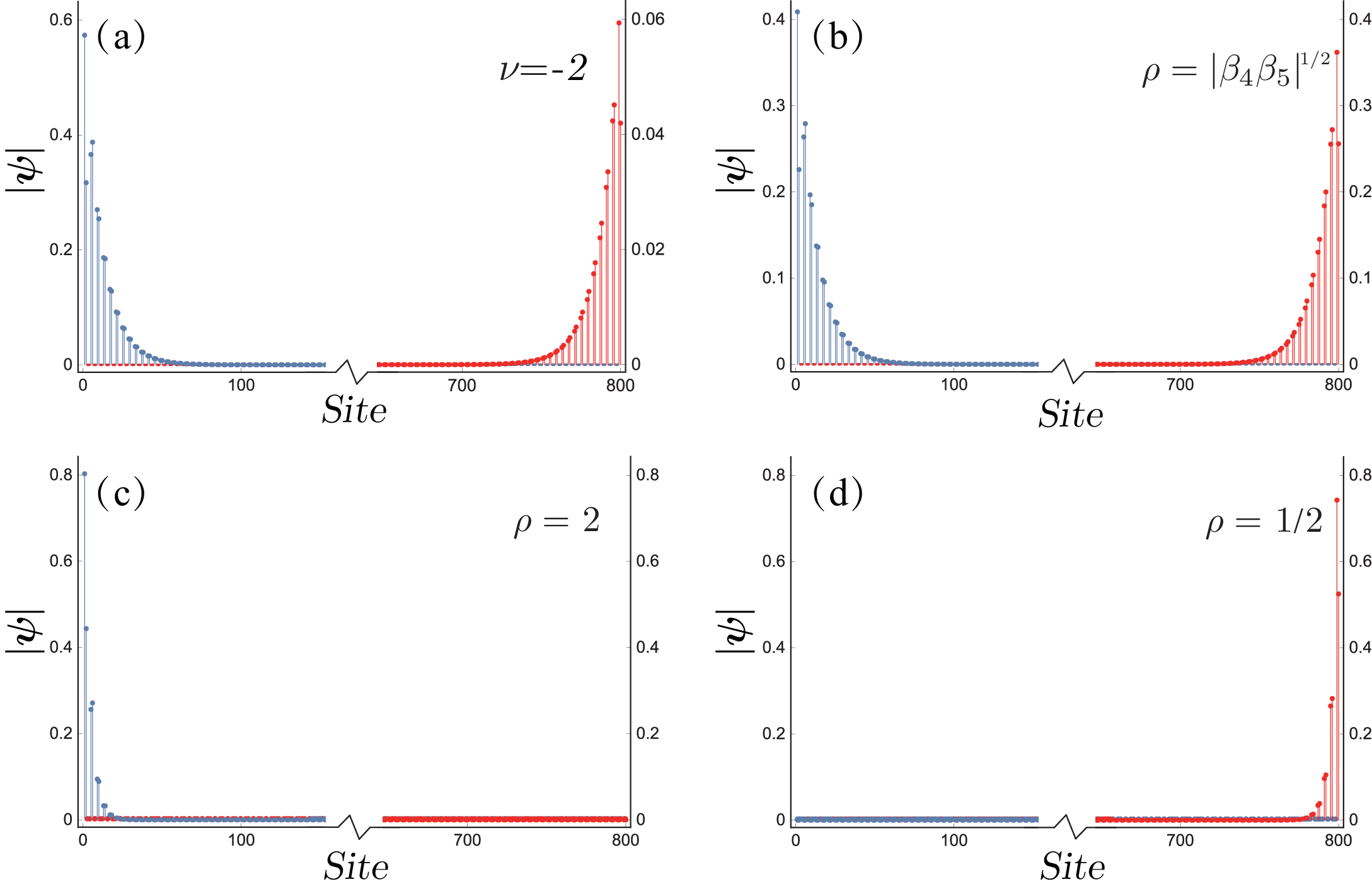}
    \caption{Space distributions of the edge states with $\nu=-2$ and $\lambda=1$. (a) One pair of the nearly overlapping zero-energy modes obtained by the numerical matrix diagonalization, and the corresponding result after a similarity transformation with (b) $\rho=\vert\beta_4\beta_5\vert^{1/2}=0.989$, (c) $\rho=2$ or (d) $\rho=1/2$, respectively. Here $L=200$.}
    \label{sfig9}
\end{figure}

Now we choose $\lambda=1$ to study the case of $\nu=-2$. The eight $\beta$ values are ordered as follows:
\begin{equation}
\vert\beta_{1}\vert=\vert\beta^{-}_{1}\vert<\vert\beta_{2}\vert=\vert\beta^{-}_{2}\vert<\vert\beta_{3}\vert=\vert\beta^{-}_{3}\vert<\vert\beta_{4}\vert=\vert\beta^{-}_{4}\vert<1<\vert\beta_{5}\vert=\vert\beta^{+}_{1}\vert<\vert\beta_{6}\vert=\vert\beta^{+}_{2}\vert<\vert\beta_{7}\vert=\vert\beta^{+}_{3}\vert<\vert\beta_{8}\vert=\vert\beta^{+}_{4}\vert.
\label{eq49}
\end{equation}
It is found that Rank$\mathcal{M_L}=$Rank$\mathcal{M_R}=2$, namely, they have been maximized so either $\mathcal{M_L}\vert \mathcal{L}\rangle=0$ or $\mathcal{M_R}\vert \mathcal{R}\rangle=0$ has two independent solutions, which can be labeled as $\vert \mathcal{L}_1 \rangle$, $\vert \mathcal{L}_2 \rangle$ and $\vert \mathcal{R}_1 \rangle$, $\vert \mathcal{R}_2 \rangle$, respectively. According to degenerate perturbation theory, let's solve  $\left(\begin{array}{cc}\mathcal{M_L} &\mathcal{M_{LR}}\\\mathcal{M_{RL}} &\mathcal{M_{R}}\end{array}\right)\vert \psi  \rangle=0$ which is equivalent to $M\vert \varphi  \rangle=0$, with $M$ being the $2\vert \nu\vert \times 2\vert \nu\vert$ matrix given by:
\begin{equation}
    \begin{split}
    M=&\left(
        \begin{array}{cccc}
         0& M_{LR} \\
         M_{RL}&0 \\
        \end{array}
        \right)\\
    =&\left(
    \begin{array}{cccc}
     0& 0& \langle \mathcal{L}_1\vert\mathcal{M_{LR}}\vert \mathcal{R}_1 \rangle& \langle \mathcal{L}_1\vert\mathcal{M_{LR}}\vert \mathcal{R}_2 \rangle \\
     0& 0& \langle \mathcal{L}_2\vert\mathcal{M_{LR}}\vert \mathcal{R}_1 \rangle&\langle \mathcal{L}_2\vert\mathcal{M_{LR}}\vert \mathcal{R}_2 \rangle \\
     \langle \mathcal{R}_1\vert\mathcal{M_{RL}}\vert \mathcal{L}_1 \rangle & \langle \mathcal{R}_1\vert\mathcal{M_{RL}}\vert \mathcal{L}_2 \rangle& 0 & 0 \\
     \langle \mathcal{R}_2\vert\mathcal{M_{RL}}\vert \mathcal{L}_1 \rangle & \langle \mathcal{R}_2\vert\mathcal{M_{RL}}\vert \mathcal{L}_2 \rangle &0 & 0 \\
    \end{array}
    \right),
    \end{split}
    \label{eq410}
\end{equation}
where $\vert \varphi  \rangle=(l_1,l_2,r_1,r_2)^T$. So $\vert \psi\rangle=(l_1\vert \mathcal{L}_1 \rangle+l_2\vert \mathcal{L}_2 \rangle, r_1\vert \mathcal{R}_1 \rangle+r_2\vert \mathcal{R}_2 \rangle)^T$. We have four solutions: $\vert \psi^{\pm}_m \rangle=(\mathcal{A}_m\vert \mathcal{L}_m^{\prime} \rangle,\pm\mathcal{B}_m\vert \mathcal{R}_m^{\prime} \rangle)^T$, where $m=1$, $2$. Here each normalized $\vert \mathcal{L^{\prime}}_m \rangle$ ($\vert \mathcal{R^{\prime}}_m \rangle$) is the linear superposition of $\vert \mathcal{L}_1 \rangle$ and $\vert \mathcal{L}_2 \rangle$($\vert \mathcal{R}_1 \rangle$ and $\vert \mathcal{R}_2 \rangle$). $\vert \psi^{+}_m \rangle$ is also connected with $\vert \psi^{-}_m \rangle$ by chiral symmetry: $\mathcal{S}\vert \psi^{\pm}_m \rangle=-\vert \psi^{\mp}_m \rangle$. From the ratio equation
\begin{equation}
    \frac{\mathcal{B}_m}{\mathcal{A}_m}=\frac{\sqrt{\langle \mathcal{R^{\prime}}_m\vert\mathcal{M_{RL}}\vert \mathcal{L^{\prime}}_m \rangle}}{\sqrt{\langle \mathcal{L^{\prime}}_m\vert\mathcal{M_{LR}}\vert \mathcal{R^{\prime}}_m \rangle}}, \quad m=1,2
    \label{eq411}
\end{equation}
we get $\vert\mathcal{B}_1/\mathcal{A}_1\vert\approx \vert\beta_4\beta_5\vert^{L/2}=\vert\beta^{-}_4\beta^{+}_1\vert^{L/2}=0.11567$ and $\vert\mathcal{B}_2/\mathcal{A}_2\vert\approx \vert\beta_3\beta_6\vert^{L/2}=\vert\beta^{-}_3\beta^{+}_2\vert^{L/2}=0.00084$, with $L=200$. The numerical result for one pair of the TPESs is shown in Fig.\ref{sfig9} (a), while the other pair(data not shown) shows analogous behavior in this case. We found that the predicted ratio is consistent very well with the numerical matrix diagonalization result. As before, if we apply the similarity transformation to this system, i.e., $H(\beta)\rightarrow H(\rho\beta)$, the ratios become $\vert\mathcal{B}_1/\mathcal{A}_1\vert\approx \vert\beta_4\beta_5/\rho^2\vert^{L/2}$ and $\vert\mathcal{B}_2/\mathcal{A}_2\vert\approx \vert\beta_3\beta_6/\rho^2\vert^{L/2}$. In Fig.\ref{sfig9} (b), we set a suitable $\rho$ to make one of the ratios close to $1$, so the corresponding pair of the TPESs would be localized at both ends of the open chain. In this situation, the total number of TPESs would be three. In Fig.\ref{sfig9} (c)-(d), we set $\rho=2$ and $\rho=1/2$ respectively. In either case, when $L$ is large, among $\mathcal{B}_m$ and $\mathcal{A}_m$, one would always be much larger than the other, and the smaller one can be viewed as be approaching zero, implying that the number of the TPESs would be equal to $\vert \nu\vert=2$.

Here we see that we can always choose a sufficiently small or large $\rho$ to perform the similarity transformation so that the number of the TPESs can be reduced to the absolute value of the topological invariant $\nu$ but is definitely not less than $\vert\nu\vert$. If all ratios $\vert\mathcal{B}_m/\mathcal{A}_m\vert$ share the same relationship, one can naturally choose a common $\rho$ to make all edge states be localized at both ends, so the number of the TPESs reaches its maximum value, $2\vert \nu \vert$. However, in most cases of non-hermitian systems, this is usually unachievable, so for a general non-hermitian open chain with chiral symmetry, the number of the TPESs would be varying from $\vert \nu \vert$ to $2\vert \nu \vert$.

\bibliographystyle{unsrt}
\bibliography{ref}